%% file: main-noline.tex
\documentclass[twocolumn]{aastex62}
\usepackage{CJK}

\newcommand{\brg}{Br$\gamma \:$}
\newcommand{\microns}{$\mu$m}

\newcommand\tna{\,\tablenotemark{a}}
\newcommand\tnb{\,\tablenotemark{b}}
\newcommand\tnc{\,\tablenotemark{c}}

\newcommand{\kms}{km s$^{-1}$}
\newcommand{\msun}{M$_{\odot}$ }

\shorttitle{S-star Binary Search}
\shortauthors{Chu et al.}

\begin{document}
\begin{CJK*}{UTF8}{gbsn}

\title{Evidence of a decreased binary fraction for massive stars within 20 milliparsecs of the supermassive black hole at the Galactic center}

\correspondingauthor{Devin Chu}
\email{dchu@astro.ucla.edu}

\author[0000-0003-3765-8001]{Devin S. Chu}
\affil{Department of Physics and Astronomy \\
UCLA\\
Los Angeles, CA 90095-1547, USA}

\author[0000-0001-9554-6062]{Tuan Do}
\affiliation{Department of Physics and Astronomy \\
UCLA\\
Los Angeles, CA 90095-1547, USA}

\author[0000-0003-3230-5055]{Andrea Ghez}
\affiliation{Department of Physics and Astronomy \\
UCLA\\
Los Angeles, CA 90095-1547, USA}

\author[0000-0002-2836-117X]{Abhimat K. Gautam}
\affiliation{Department of Physics and Astronomy \\
UCLA\\
Los Angeles, CA 90095-1547, USA}

\author[0000-0001-5800-3093]{Anna Ciurlo}
\affiliation{Department of Physics and Astronomy \\
UCLA\\
Los Angeles, CA 90095-1547, USA}

\author[0000-0003-2400-7322]{Kelly Kosmo O'neil}
\affiliation{Department of Physics and Astronomy \\
UCLA\\
Los Angeles, CA 90095-1547, USA}

\author[0000-0003-2874-1196]{Matthew W. Hosek Jr.}
\altaffiliation{Brinson Prize Fellow}
\affiliation{Department of Physics and Astronomy \\
UCLA\\
Los Angeles, CA 90095-1547, USA}

\author[0000-0002-2186-644X]{Aur\'elien Hees}
\affiliation{SYRTE \\
Observatoire de Paris\\
Universit\'e PSL, CNRS, Sorbonne Universit\'e, \\
Paris, France}

\author[0000-0002-9802-9279]{Smadar Naoz}
\affiliation{Department of Physics and Astronomy \\
UCLA\\
Los Angeles, CA 90095-1547, USA}
\affiliation{Mani L. Bhaumik Institute for Theoretical Physics \\
Department of Physics and Astronomy \\
UCLA\\
Los Angeles, CA 90095-1547, USA}

\author[0000-0001-5972-663X]{Shoko Sakai}
\affiliation{Department of Physics and Astronomy \\
UCLA\\
Los Angeles, CA 90095-1547, USA}

\author[0000-0001-9611-0009]{Jessica R. Lu}
\affiliation{Astronomy Department\\
University of California, Berkeley\\
Berkeley, CA 94720, USA}

\author[0000-0002-3038-3896]{Zhuo Chen (陈卓)}
\affiliation{Department of Physics and Astronomy \\
UCLA\\
Los Angeles, CA 90095-1547, USA}

\author[0000-0001-7017-8582]{Rory O. Bentley}
\affiliation{Department of Physics and Astronomy \\
UCLA\\
Los Angeles, CA 90095-1547, USA}


\author{Eric E. Becklin}
\affiliation{Department of Physics and Astronomy \\
UCLA\\
Los Angeles, CA 90095-1547, USA}

\author{Keith Matthews}
\affiliation{Department of Physics and Astronomy \\
Caltech\\
Pasadena, CA 91125, USA}



\begin{abstract}

We present the results of the first systematic search for spectroscopic binaries within the central 2 x 3 arcsec$^2$ around the supermassive black hole at the center of the Milky Way galaxy. This survey is based primarily on over a decade of adaptive optics-fed integral-field spectroscopy (R$\sim$4000), obtained as part of the \textit{Galactic Center Orbits Initiative} at Keck Observatory, and has a limiting $K$'-band magnitude of 15.8, which is at least 4 magnitudes deeper than previous spectroscopic searches for binaries at larger radii within the central nuclear star cluster. From this primary dataset, over 600 new radial velocities are extracted and reported, increasing by a factor of 3 the number of such measurements.  We find no significant periodic signals in our sample of 28 stars, of which 16 are massive, young (main-sequence B) stars and 12 are low-mass, old (M and K giant) stars. Using Monte Carlo simulations, we derive upper limits on the intrinsic binary star fraction for the young star population at 47\% (at 95\% confidence) located $\sim$20 mpc from the black hole. The young star binary fraction is significantly lower than that observed in the field (70\%). This result is consistent with a scenario in which the central supermassive black hole drives nearby stellar binaries to merge or be disrupted and may have important implications for the production of gravitational waves and hypervelocity stars.





\end{abstract}

\keywords{Infrared spectroscopy (2285), Galactic center (565), Adaptive optics (2281)}




\section{Introduction} \label{sec:intro}

The closest known stars to the Milky Way's supermassive black hole (SMBH) comprise the so-called ``S-star'' cluster, where S stands for Sgr A*, the emissive source associated with the SMBH. This population is both distinct dynamically and spectroscopically from the surrounding stellar population. Spectroscopic observations have also revealed that most of these stars are main-sequence B stars \citep{Ghez:2003iw,Eisenhauer:2005gh,Habibi:2017}. Unlike their cousins outside the central radius of 0.04 parsecs, this population lacks Wolf-Rayet stars, suggesting the S-stars have formed within the last 20 million years. Their young ages raise questions about their formation mechanism, since traditional star formation would be disrupted by the tidal forces of the black hole \citep{Morris:1993fp}.


Numerous investigations have been done to postulate the formation of these S-stars. General mechanisms include: (1) binary star systems scattered from outside the region and then tidally disrupted, leaving behind one component of the original binary while the other is ejected as a hypervelocity star \citep[e.g.][]{Hills:1988br,Perets:2007fo,Generozov2020ApJ}, (2) S-stars formed in the clockwise disk located just outside 1 arcsec of the SMBH and then migrated to the SMBH \citep[e.g.][]{Levin:2007ez,Lockmann:2008be,Merritt:2009ab}, and (3) merger of binary stars at the Galactic centers caused by the Kozai-Lidov mechanism, with the product appearing as a main-sequence B-star \citep[e.g.][]{witzel2014,Stephan:2016eh,Fragione2019MNRAS,Ciurlo:2020Nature}.

Binary stars provide crucial roles in these formation mechanisms, and the discovery of binary stars amongst the S-stars may attest to particular formation mechanisms. Additionally, massive stars in the field have high multiplicity fractions \citep{Sana:2012,Duchene:2013}, so it is reasonable to expect these main-sequence B-stars at least started out in multiple systems \citep{Naoz2018}. Previous studies have identified 3 binary systems \citep{Ott:1999gn,Martins:2006,Rafelski:2007gu,Pfuhl:2014eba,Gautam:2019}, but none amongst the S-stars. \citet{Chu:2018} performed the first spectroscopic search for binaries amongst the S-stars and focused on the well-studied star S0-2 (also known as S2). Through radial velocity monitoring, \citet{Chu:2018} did not find significant evidence for S0-2 being a binary and placed a hypothetical companion mass upper limit at 1.6 \msun, which is below current detection limits. The dataset that was used to analyze S0-2 can be used to perform a more comprehensive survey.



In this work, we use the Galactic Center Orbits Initiative (GCOI, PI Ghez, W. M. Keck Observatory 1995 - present) long-term monitoring of this region with W. M. Keck Observatory to conduct a systematic search for binary stars using radial velocities of the S-stars. This paper is organized as follows. Section \ref{sec:Sample} details the sample selection process for this search. Section \ref{sec:data} describes the radial velocity data used in searching for spectroscopic binaries. Section \ref{sec:motion} details the process of modeling the long-term motion of the sample stars around the central black hole. Section \ref{sec:per_search} describes the search methodology for looking for companion stars in the stellar sample. Section \ref{sec:derive_limits} provides an overview of placing an upper limit on the binary star fraction. Section \ref{sec:discussion} discusses how these limits carry implications for the evolution of the S-stars.


\section{Sample Selection} \label{sec:Sample}

The broadest criteria of the star sample used in the analysis presented in this paper is that the star must be brighter than $K'$ $>$ 16 magnitude and located within the field of view of this study's primary dataset, which is centered on S0-2 and covers 3" $\times$ 2" at a PA of 285 degrees (see Figure \ref{fig:S-star_OSIRIS}, Table \ref{tab:all_spec_obs}, and Section \ref{sec:data}). Our magnitude limit stems from what can be measured with adequate signal-to-noise from a single night of observations ($\sim$3-4 hours of integration). These initial criteria yield an intermediate sample of 62 stars . From here, we make several other cuts. First,  Wolf-Rayet emission line sources (IRS16C and IRS16SW) are excluded, because measuring their radial velocities is complicated due to their stellar winds. Similarly, the seven main-sequence O stars are not included since they are featureless across the spectral range studied (2.121 -- 2.229 $\mu$m). We further omit 25 stars, for which source confusion prevents their radial velocities from being extracted without bias in our primary dataset (see Appendix \ref{sect:confusion_gas_appendix} for details). We also omit the star S0-28 because it only has 2 radial velocity measurements, which is too few to conduct a periodicity search. This leads to a final sample 28 stars, of which 16 are early-type stars and 12 are late-type stars (see Table \ref{tab:full_rv_sample}).



\input{all_spec_table_v2_09-28-22}

\input{full_sample_03-13-23}






\section{Radial Velocities} \label{sec:data}

\subsection{New Radial Velocities} \label{subsec:datasets}



\begin{figure*}
\centering
\includegraphics[width=\linewidth]{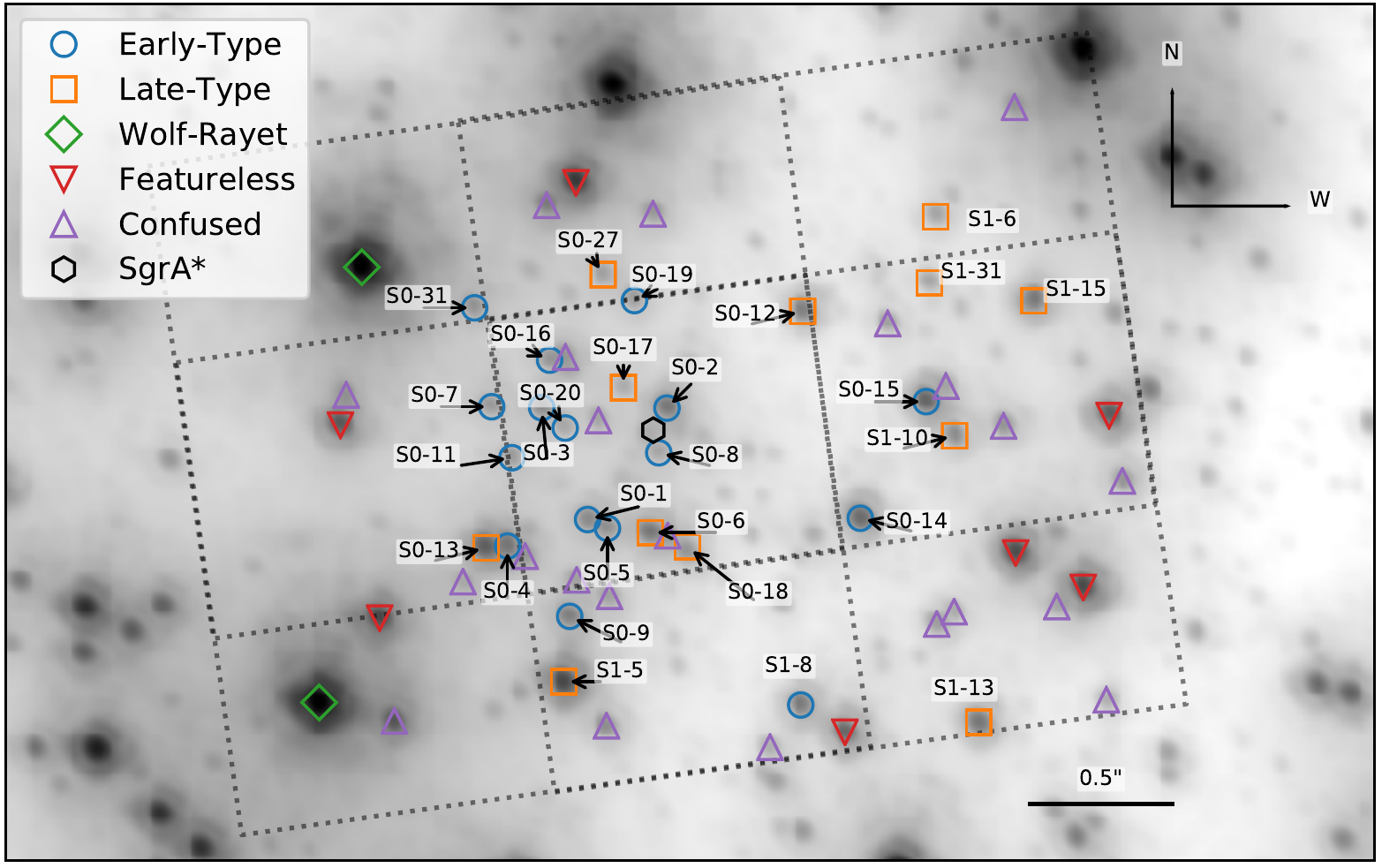}
\caption{
Finding chart for stars included in the periodicity search. The GCOI's OSIRIS 4-pointing outer dither pattern is overlaid on an adaptive optics 2.2 $\mu m$ Keck NIRC2 image of the Galactic center. Most of the sample is located in the central pointing dither pattern (blue circles: early-type, orange squars: late-type). The central square is covered by all dither points, meaning more spectra are taken of these stars throughout the night. We include all stars brighter than $K = 16$ mag, except Wolf-Rayet stars (green diamond), stars lacking absorption lines (red triangle), or are confused with other stellar sources or gas features (purple triangle). 
\label{fig:S-star_OSIRIS}
}
\end{figure*}

The primary starting point for the radial velocity analysis is the spectrally calibrated datasets that have been used by the Galactic Center Orbits Initiative (GCOI) to study S0-2 (see references in Table \ref{tab:all_spec_obs}). The majority of these observations were taken with the OSIRIS spectrograph \citep[$R \sim$ 4000,][]{Larkin:2006} on the W. M. Keck 10-m Telescope using the laser guide star adaptive optics (LGSAO) system \citep{Wizinowich:2006,vanDam:2006} and reduced via the OSIRIS reduction pipeline \citep{OSIRIS_pipeline:2017, Lockart:2019}. These 45 datasets were taken in the 35mas pixel scale and through the Kn3 (2.121 -- 2.229 $\mu$m) filter, which covers the \brg absorption line ($\lambda =$ 2.166 \microns) for the young stars and Na for the old stars. With a 2 x 2 dither pattern that keeps the central 1" $\times$ 1" in the field of view, a total view of 3" $\times$ 2" is achieved (see Figure \ref{fig:S-star_OSIRIS}). Over the course of the reported observations, OSIRIS has gone through the following two upgrades: 1) a grating upgrade in December 2012 \citep{Mieda2014PASP} and 2) a detector upgrade in April 2016 \citep{Boehle2016SPIE}. Appendix \ref{sect:instrument} shows the impact of the detector upgrade on our dataset (the grating upgrade had no significant effect).
Of the 45 datasets with newly reported radial velocities, four observations are newly reported here: 05-17-2014, 05-19-2014, 05-22-2014, and 08-07-2015 UT. The three nights in May 2014 do not contain S0-2 radial velocity measurements due to noise spikes affecting the spectra of that star. For the night of 08-07-2015, a field south of the central pointing was observed, which meant S0-2 was not in the field of view, but the star S1-13 was in the dither, which is on the southern edge of the central pointing. On average, the observations have a FWHM of 76 mas and a spectral signal-to-noise ratio of 53 for a 14 mag star.

Several datasets supplement the above core data. Thirteen of these datasets are also taken with OSIRIS, at a different plate scale (20 mas) and/or through the broader Kbb filter (1.965 -- 2.381 $\mu$m). Of these, two are new observations that have not been previously published. These two Kbb observations were taken on 07-11-2016 and 07-12-2016 UT. On these nights, weather and dithering problems prevented S0-2 from being observed, but other stars in the sample were still in the field of view. We also include a number of previously published datasets taken with other instruments, including Keck NIRSPEC ($R \sim$ 2800 in low resolution mode), Keck NIRC2 ($R \sim$ 4000), Gemini North NIFS ($R \sim$ 5000), and Subaru IRCS ($R \sim$ 20000). A summary of all the spectroscopic observations with new radial velocities is reported in Table \ref{tab:all_spec_obs}.

\subsection{Extracting Radial Velocities} \label{subsec:rvs}

Previous papers from the UCLA GCOI reported the radial velocities of S0-2 and S0-38. In this work, we also extracted the radial velocities of other known stars located in the central pointing. The methods and calibrations used to measure radial velocities are reported in \citet{Do:2019}. 

\begin{figure}
\centering
\includegraphics[width=\linewidth]{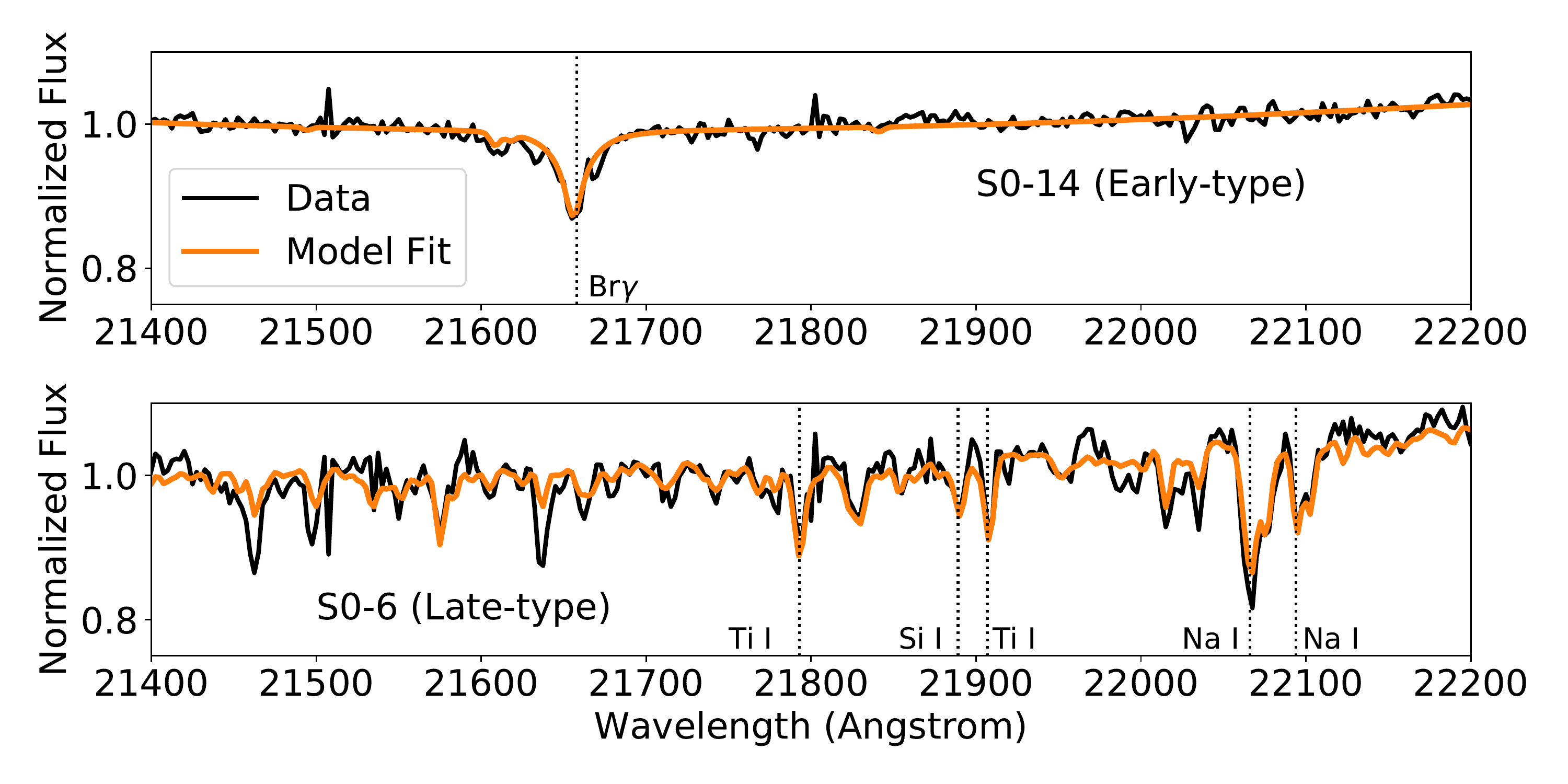}
\caption{
Example spectra and model fits for the early-type star S0-14 (top) and the late-type star S0-6 (bottom). \brg absorption is the primary spectral feature for early-type stars in the OSIRIS Kn3 filter, while the late-type stars have more features. The multiple absorption line features in late-type stars lead to increased precision for their radial velocity measurements compared to early-type stars.
\label{fig:example_spec}
}
\end{figure}

To summarize, a star's spectrum is extracted from the individual data cubes from a given epoch using a circular aperture, with an annulus around the star to estimate the sky background. The spectra are then averaged into a combined spectrum. The star's combined spectrum is then modeled using the Bayesian inference tool \textit{Starkit} \citep{starkit:2015} and compared to spectra in the BOSZ spectral grid \citep{BOSZ:2017}. The radial velocity and its uncertainty are derived using the median and 1 sigma central credible interval of the marginalized posterior. \cite{Do:2019} showed that this technique of spectral fitting reduced uncertainties and systematic bias compared to fitting a Gaussian to the \brg line for the star S0-2. Example spectra and their model fits for both early and late-type stars are shown in Figure \ref{fig:example_spec}. Early-type stars are main-sequence B stars, with the \brg absorption line being the major spectral feature in the Kn3 filter. Late-type stars are M and K giants with many absorption lines in Kn3, noteably the Na doublet lines around 22100 angstroms. The measured radial velocity is then corrected for the local standard of rest with respect to the Galactic center\footnote{We use the IRAF procedure \textit{rvcorrect}. This correction uses a velocity of 20 \kms \ for the solar motion with respect to the local standard of rest in the direction $\alpha = 18^{h}, \delta = +30\deg$ for epoch 1900 \cite{Kerr:1986}, corresponding to $(u,v,w) = (10, 15.4, 7.8)$ \kms.}.

We apply this same technique to all the other stars in the OSIRIS data when a star's radial velocity can be measured. The number and quality of radial velocity measurements extracted for each epoch depends greatly on a number of factors, such as weather conditions, adaptive optics performance, and position in the OSIRIS dither pattern. Stellar crowding and confusion can also lead to difficulties when extracting a radial velocity measurement. Even though a star may be identified in an OSIRIS cube, its spectrum may not be of adequate quality to measure its radial velocity. We perform a quality inspection of the extracted spectra to ensure their radial velocities can be measured.

\begin{figure}
\centering
\includegraphics[width=\linewidth]{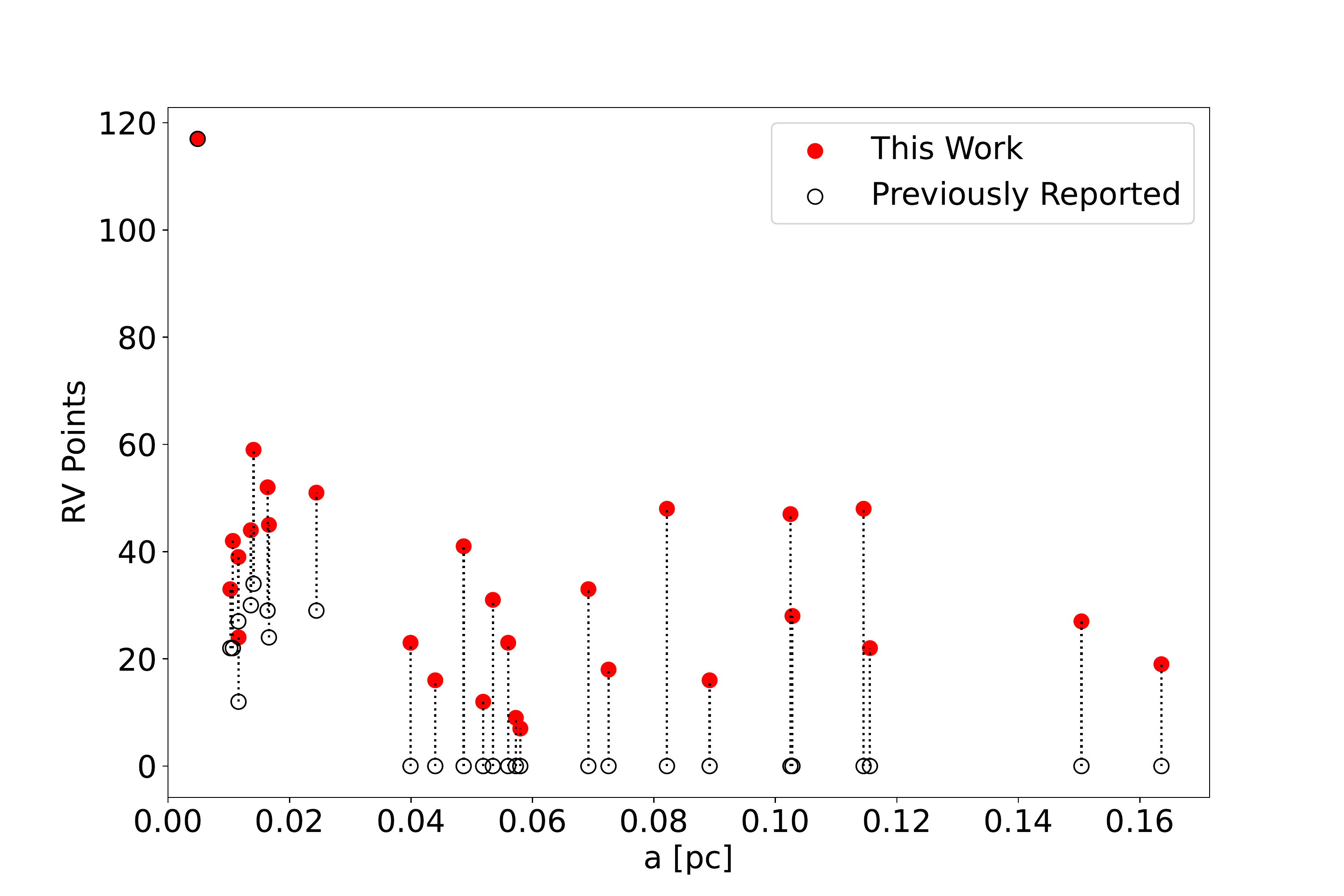}
\caption{
Number of radial velocity points for a star versus its estimated semi-major axis. The number of previously published points is given by the unfilled gray circles, while the number of points used in this work is given in red. The dotted lines connect the points for visual increase. S0-2 is the star with the most radial velocity points because of its brightness and close proximity to the black hole. This work reports 626 new radial velocity measurements.
\label{fig:rv_semimajor}
}
\end{figure}

In this work, we report 626 new radial velocity measurements. To this, we add the 344 radial velocities from the literature \citep{Gillessen:2017fa, Do:2019}, the majority of which are for S0-2. As Figure \ref{fig:rv_semimajor} shows, the new RV measurements dramatically increase ($\sim$3 times) the coverage for other stars in this region, enabling the first binary star population study.



\section{Modeling the long-term motion around the SMBH} \label{sec:motion}


Some of the stars in the selected sample have significant long-term motion from their orbits around the SMBH, which we model and remove as a necessary initial step for conducting a periodic search for binary stars. For the 10 stars with orbital periods around the SMBH of less than 180 years (a semi-major axis less than 24 mpc) and which have gone through a turning point during our observations, we have enough astrometry and radial velocities in the GCOI database to directly model their orbital motions. For the shortest period and best-studied star, S0-2, we use the reported orbital model in \citet{Do:2019}. This model also provides us with the black hole parameters (position, proper motion, mass, and distance to Earth), which we use as fixed values in our orbital model for the other short-period stars. For the short-period stars beyond S0-2 with measurable orbits, we also fix the astrometric correlation length from source confusion to 30 mas and the radial velocity offset between the Keck and VLT measurements to 0 \kms \citep[see][]{Do:2019,Ciurlo:2020Nature}. Leaving these values free has no impact on modeling the radial velocity curves or residuals. The seven modeled parameters for these stars are the six standard stellar orbital parameters (period, eccentricity, inclination, longitude of the ascending node, argument of periapse, epoch of closest approach) plus the astrometric mixing parameter. The star's semi-major axis and its uncertainty can also be obtained from its orbital period (and uncertainty) and black hole mass using Kepler's Third Law. Once the star's orbital fit is performed, a model for its radial velocity is generated and subtracted to create a residual curve. The last column of Table \ref{tab:full_rv_sample} provides the estimated semi-major axes for these short-period stars' motions around the SMBH.

For stars with longer orbital periods around the black hole, we performed a polynomial fit to their radial velocities. The degree of the polynomial fit is determined by the F-test, where a higher degree polynomial must pass the F-test of the lower degree polynomial with a 95\% significance. Of the 18 stars fit with a polynomial, all but one are best fit with a constant radial velocity model, and one (S0-6) is well fit with a constant acceleration model. These polynomial fits are reported in Table \ref{tab:polynomial_table_test}. The polynomial fit is then subtracted from the radial velocity points to create residual points, which are reported in Table \ref{tab:S0-14_example_rv}. For these longer period stars, their semi-major axes are estimated using the same orbit fitting method described above. These estimates have formal uncertainties from 1-20\%, and the lower value is most likely an underestimate due to the small orbital phase coverage; we therefore assign a 20\% semi-major axis uncertainty for these stars. For three late-type stars on the edge of our sample, we were unable to obtain orbital solutions. To estimate their semi-major axes, we take the average of comparably large separations from the SMBH. These semi-major axes estimates are also reported in Table \ref{tab:full_rv_sample}.

\input{polynomial_table_09-30-22}

\input{S0-14_rvtable_example}



\begin{figure}
\centering
\includegraphics[width=\linewidth]{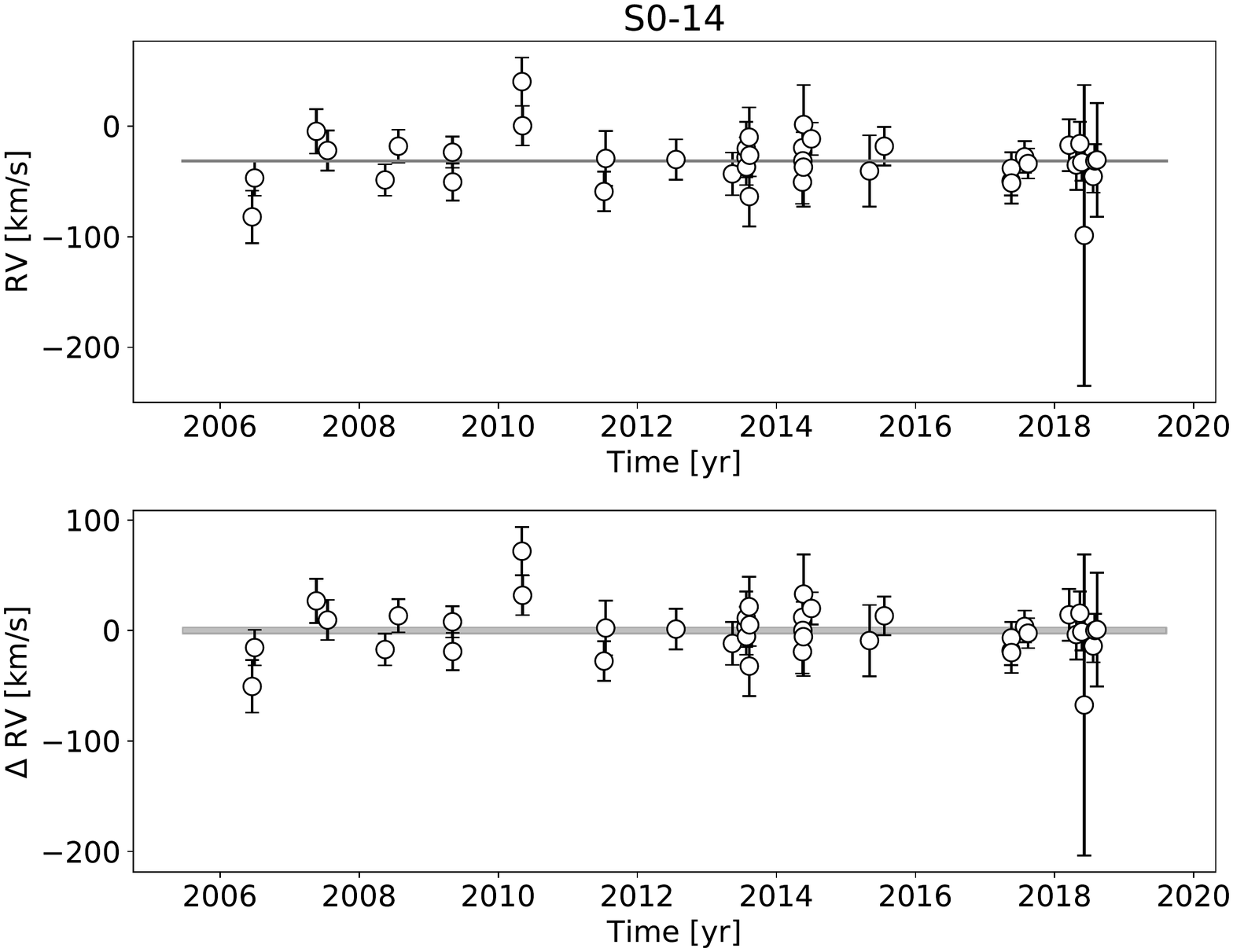}
\caption{
Top: Radial velocity of S0-14 over time. A model fit of polynomial order 0 is fit to the data. Bottom: Difference of measured radial velocity points from the model for S0-14. Uncertainty of the model fit is incorporated into the shaded region.
\label{fig:S0-14_rv_res}
}
\end{figure}




\section{Companion Star Searches} \label{sec:per_search}

The following section describes the process for detecting periodic signals in the radial velocity residual curves using two types of periodicity searches: a Lomb-Scargle analysis \citep{Lomb:1976,Scargle:1982eu,VanderPlas:2018} and a Bayesian fit for potential binary systems. A similar methodology was done for the star S0-2 in \citet{Chu:2018}. The Lomb-Scargle analysis provides a computationally efficient method for detecting periodic signals in unevenly spaced data. The Bayesian fitting method provides a more complete and robust approach and allows us to derive upper limits on the orbital parameters of hypothetical binary companions to these S-stars.


\subsection{Lomb-Scargle Analysis} \label{subsec:lombscargle}

Once a star's residual radial velocity curve is made, it is run through a Lomb-Scargle package, gatspy \citep{VanderPlas:2015,gatspy:2016} to search for a periodic signal. When running this periodic search, a range of periods from 2 to 10000 days are sampled. The lower period sampling limit of 2 day comes from the Nyquist sampling limit of taking data on consecutive days. The upper period limit of 10000 days ensures that the entire time baseline of the dataset is covered. To ensure that potential signals are not missed due to uniform sampling, the ``$N$ samples per peak''feature of the gatspy package is used. We specify $N=10$ samples per peak, which carries out 10 additional, finer samples around a peak in Lomb-Scargle power \citep[see][for details]{VanderPlas:2015,gatspy:2016}. The periods are uniformly spaced at $1/N\Delta t$, where $N$ is the samples per peak and $\Delta t$ is the maximum time baseline of observations. Spacing the sampled periods at $1/N\Delta t$ ensures proper sampling of a dataset \citep{VanderPlas:2018}. \citet{Chu:2018} calculated an upper limit of 119 days for the longest period for an S0-2 binary star system, as systems with longer periods would separated at S0-2's closest approach to the SMBH. The stars in this sample are not expected to pass as close to the SMBH, hence would have longer maximum periods, which is why we decided to increase the upper period sampled. After this step, a Lomb-Scargle power spectrum is obtained, containing power values for every sampled period. An example Lomb-Scargle periodogram for S0-14 is shown in Figure \ref{fig:LS_S0-14}. Periodograms for the entire sample are given in Appendix \ref{sect:ls_k_results}

The significance of the peak Lomb-Scargle powers are determined in three ways: i) comparing the Lomg-Scargle amplitude to the star's radial velocity uncertainty ii) Monte Carlo simulation and iii) bootstrap false alarm probability (FAP) test significance. Each of these methods produces a significance value between 0 to 100\%, with the higher percentage corresponding to a higher significance. These three methods compliment one another, and a higher significance value means a higher likelihood that a binary system as been detected.

One way to evaluate the significance of the Lomb-Scarlge power for each star is to look at the fit amplitude from the Lomb-Scargle model. The Lomb-Scargle analysis returns an amplitude of the sinusoid model fit to the residual curve. This amplitude is divided by the median radial velocity uncertainty for that star to determine the amplitude significance in terms of sigma. If a star's residual has a very large amplitude of variation relative to its radial velocity uncertainty, the variation can be considered significant relative to noise.

The second way to determine significance is to use a Monte Carlo simulation. This Monte Carlo simulation is conducted in the same way as described in \citet{Chu:2018}. To summarize, 100,000 simulated residual curves with no periodic signal are generated for each star. Each simulated curve has the same observation times and uncertainties as the data. Every data point is drawn from a Gaussian distribution centered around 0 \kms, meaning the simulated data contained no periodic signal and only noise. These simulated curves are then run through the Lomb-Scargle process described above, and the maximum Lomb-Scargle power for each run is recorded. A cumulative distribution function for the 100,000 simulations is compiled. The peak Lomb-Scargle power from the data is then compared to the cumulative distribution function. This approach allows us to quantify the significance of our peak signal relative to a non-periodic data set taken at the same observation sampling and uncertainties.


\begin{figure}
\centering
\includegraphics[width=\linewidth]{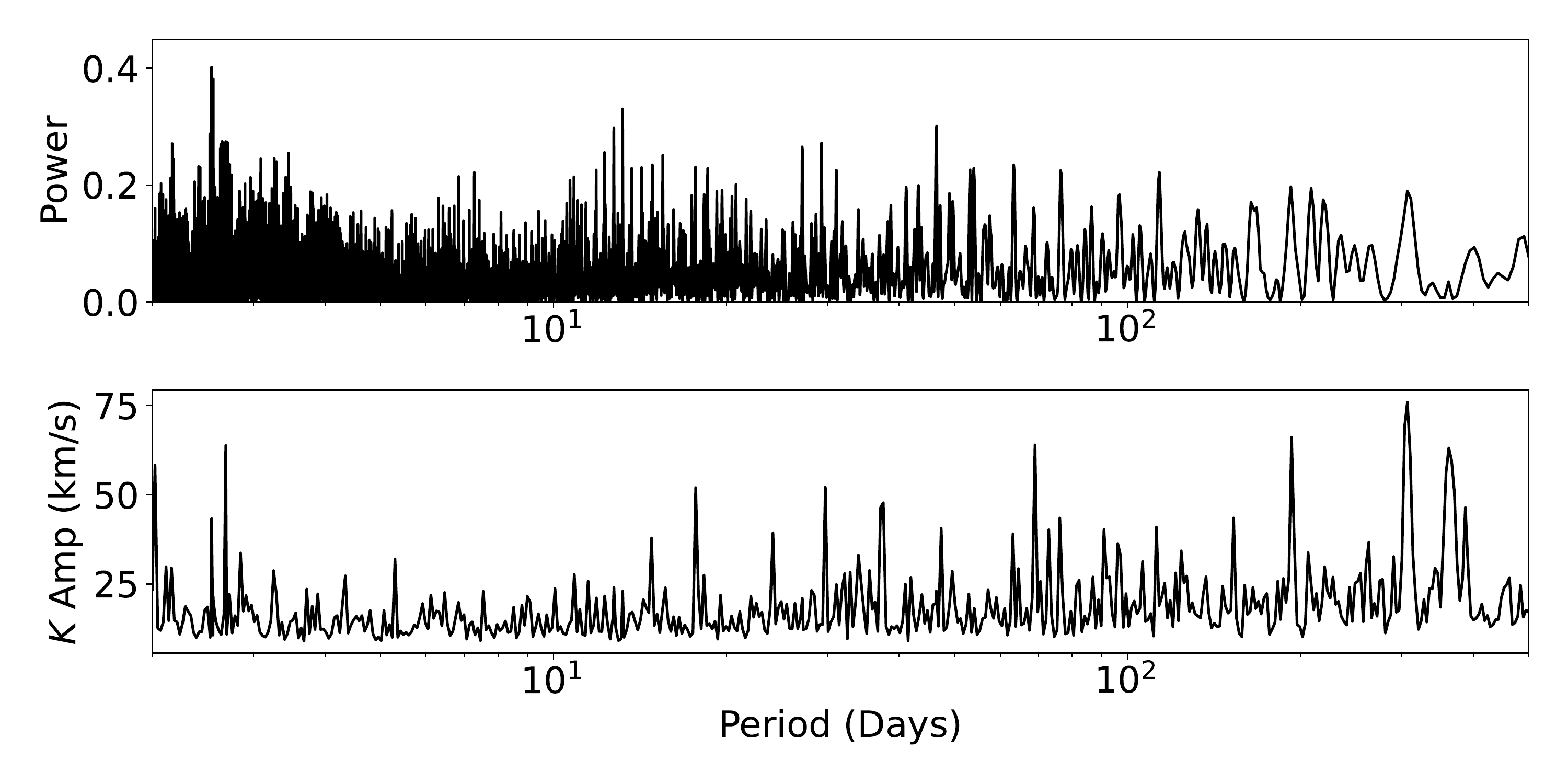}
\caption{
Top: Lomb-Scargle periodogram of S0-14's residual radial velocity curve. Every period value sample has a corresponding Lomb-Scargle power. The significance of the peak power value is then calculated to determine if this period represents a detection. Bottom: 95\% upper confidence limit on the amplitude of RV variations induced by a binary system ($K$) as a function of the binary orbital period for S0-14.
\label{fig:LS_S0-14}
}
\end{figure}



The bootstrap false alarm probability (FAP) is another way to test the significance of the signal in a slightly different view than the Monte Carlo simulations. \citet{VanderPlas:2018} explains how the FAP addresses the probability that a signal with no periodic component would lead to a peak of a given magnitude. We choose the bootstrap algorithm method in gatspy because it is the most robust estimate specifically of the FAP compared to other given methods in gatspy \citep[see][for details]{VanderPlas:2018}. A similar bootstrap method was implemented by \citet{Gautam:2019} to determine the FAP with Galactic center photometry data. To conduct this test, 10,000 Lomb-Scargle periodograms are simulated. Each of these periodograms is obtained by keeping the observation times and by drawing residual values randomly with replacement from the residual curve. The maxima of each resulting periodogram are computed. The peak periodogram power values from the data are compared to the distribution of power values from the bootstrap to determine the FAP. The bootstrap false alarm test significance is then defined as 1 - FAP and reported. This way, a higher value of 1 - FAP (a lower FAP value) corresponds to a higher significance of a binary.

None of the stars' residual curves have periodic variations beyond the 3$\sigma$ (99.7\%) confidence limit using all three methods. While there is some variation in the significance values for some stars, the main importance is how the star performs in all three tests. A true binary system is expected to show high significance with all three tests. No star exhibits high significance across all methods, suggesting that no binary stars have been detected. The detailed results from the Lomb-Scargle periodicity search are presented in Table \ref{tab:periodicity_results}. 



\input{periodicity_result_withk}


\subsection{Binary Curve Fitting} \label{subsec:binary_fit}

Another approach to search for a companion is through a Bayesian fit of the residual curve to the binary system curve. This is the same as the method described in \citet{Chu:2018}. The residual curves are fit with a binary star radial velocity model plus a constant. The following equation was used to model the radial velocity curve of an eccentric binary system~\citep{Hilditch:2001aa}
\begin{equation}
	RV = K \frac{\sqrt{1 - e^{2}}\cos{E}\cos{\omega} - \sin{E}\sin{\omega}}{1 - e \cos{E}}\ + O,
	\label{eq:eccentric_rv}
\end{equation}
with
\begin{equation}
	K = \frac{2\pi a\sin{i}}{P}\, ,
	\label{eq:Kdef}
\end{equation}
and where $e$ is the binary eccentricity, $\omega$ the argument of periastron, $E$ the eccentric anomaly determined by solving the Kepler equation, $i$ the inclination, $P$ the period  and $a$ the semi-major axis. This model is parametrized using the following 5 variables: the constant offset $O$, the radial velocity amplitude $K$, the eccentricity $e$, the argument of periastron $\omega$ and the mean longitude at J2000 (noted $L_0$). The use of the mean longitude at J2000 is preferred to the usual time of closest approach which is not bounded and not defined in case of circular orbits~\citep{Hilditch:2001aa}. For different fixed binary orbital periods $P$, this model is fit to the radial velocity residuals using a MultiNest sampler~\citep{Feroz:2008fi,Feroz:2009aa,Feroz:2013aa}. A strong periodic signal at a given period would lead to a large, peaked value of $K$ in the posterior. This method takes into account parameters such as eccentricity, which changes the shape of the curve from a perfect sinusoid wave. Periods from 2 to 500 days are uniformly sampled in log space. For S0-2, we followed the same methodology as \citet{Chu:2018}, where we evenly spaced at 0.05 days for periods from 2 to 150 days since periods beyond 119 days are excluded by the binary stability criteria. Because of the more computationally expensive nature of this method, we did not sample periods as long as the Lomb-Scargle method. An example output of this methodology is shown for S0-14 in Figure \ref{fig:LS_S0-14} and the complete set of the $K$ amplitude limit figures for the full sample is provided in Appendix \ref{sect:ls_k_results}.



After calculating $K$ upper limits for every sampled period, the median $K$ value is then taken as a summary upper limit for the star. The limits on $K$ amplitude are reported in Table \ref{tab:periodicity_results}. The $K$ amplitude results for the sample are also shown in Figure \ref{fig:LS_S0-14}. The $K$ amplitude limits can be used to derive hypothetical companion mass limits and are reported in Appendix \ref{sec:comp_mass_limits}. We do not report any detection of a binary system from this method, and these limits reflect our sensitivities to detecting binaries.

\begin{figure}
\centering
\includegraphics[width=\linewidth]{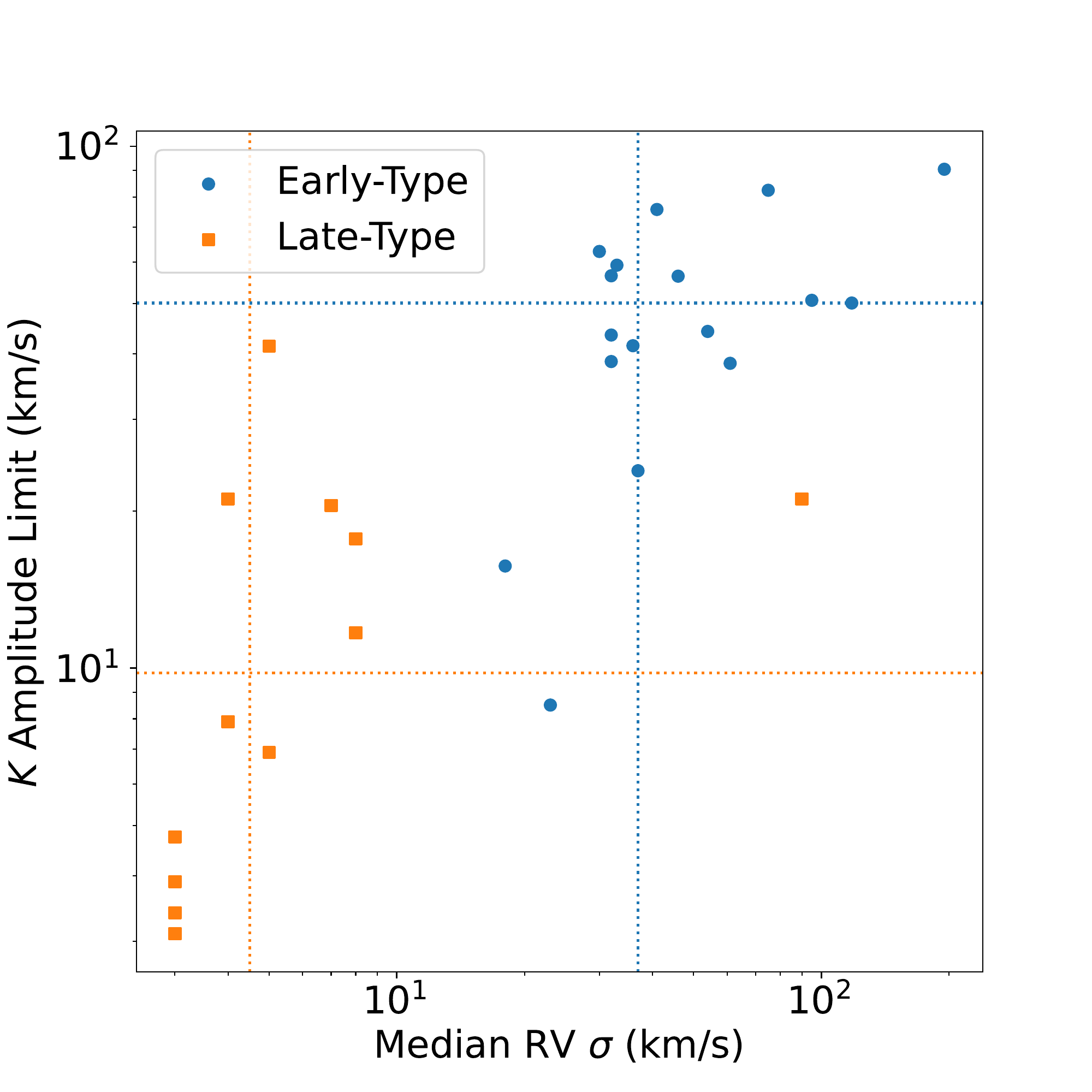}
\caption{
Median radial velocity uncertainty plotted with the $K$ amplitude limit value for each star, color coded by their spectral types. The $K$ amplitude value comes from marginalizing the $K$ amplitude limits over all sampled periods. The dashed lines represent the median values of the limits and radial velocity uncertainties for both the early-type and late-type stars.
\label{fig:Kamp_RVsig}
}
\end{figure}


\section{Binary Star Fraction Limits} \label{sec:derive_limits}




Performing this systematic search for spectroscopic binaries has yielded no candidates, and we can use this result to place limits on the intrinsic binary population. To do this, one needs to make assumptions about the underlying binary star population. For the young, massive stars, we make use of the \citet{Sana:2012} distributions of binary system parameters (mass ratios $q$, eccentricities, periods). For the late-type stars, which are expected to be around 1\msun, we pull from the distributions reported by \citet{Raghavan:2010}. These distributions are used to create an initial estimate of the $K$ amplitude distributions for both the massive star and solar mass star binary populations and we later explore variations in Appendix \ref{sec:period_simulations}, which shows no impact for the early-type stars and a very modest impact for the late-type stars. 

\begin{figure}
\centering
\includegraphics[width=\linewidth]{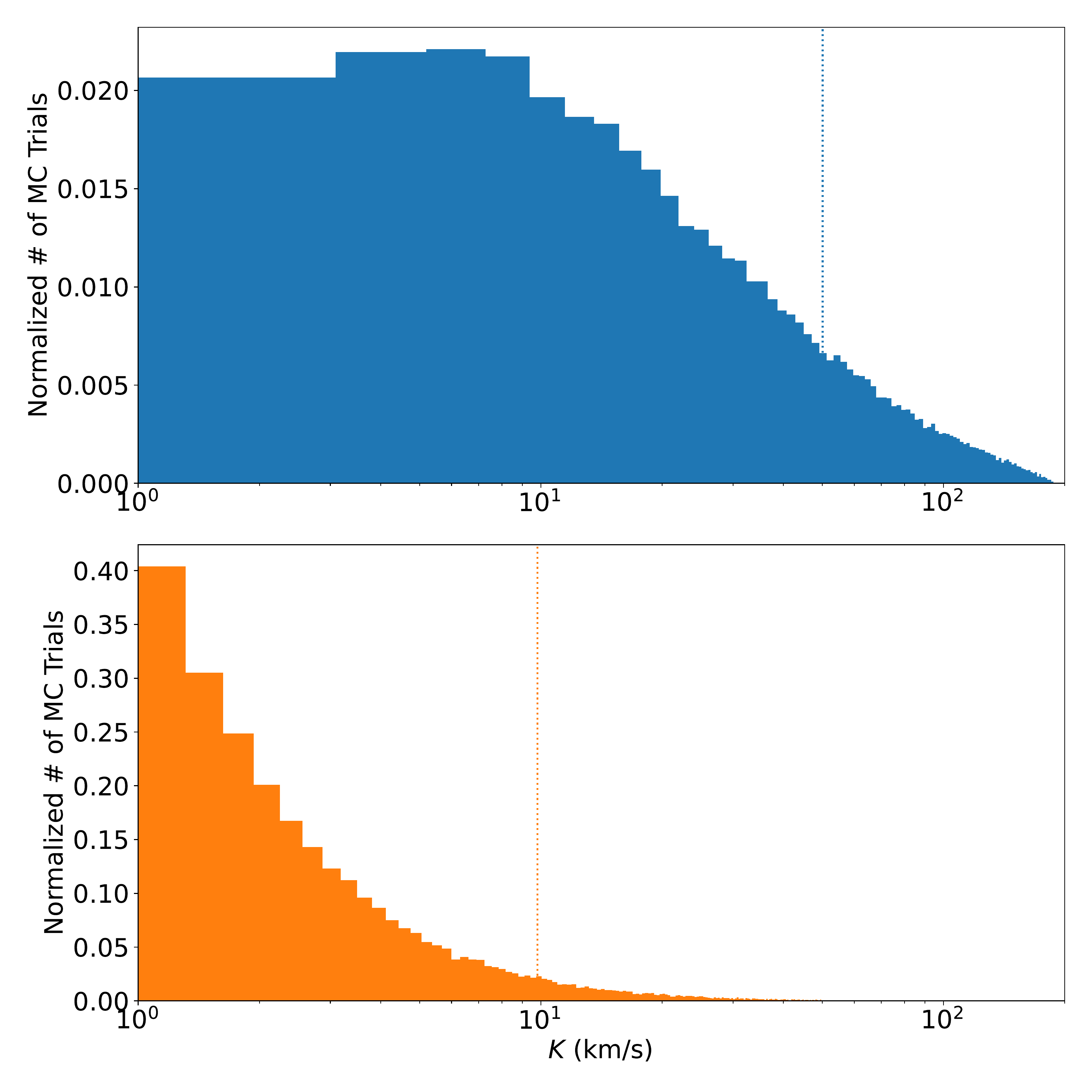}
\caption{
Top: Normalized $K$ amplitude distributions for the simulated massive star binary population using \citet{Sana:2012} parameters, along with the median $K$ amplitude limit from the early-type sample (blue dotted line). Bottom: Normalized $K$ amplitude distributions for the simulated solar mass binary population using \citet{Raghavan:2010} parameters, along with the median $K$ amplitude limit from the late-type sample (orange dotted line). While the late-type sample has smaller $K$ amplitude limits, they remain higher than the distribution of $K$ amplitudes for the corresponding simulated binary star population.
\label{fig:simulated_k_dist}
}
\end{figure}

Parameters are drawn from the given distributions of log P, $e$, and mass ratio $q$ from the given distributions. This is done 100,000 times to create a population of 100,000 binary systems. Using the binary mass equation:

\begin{equation}
	M_{\text{comp}}\sin{i}  = \left(\frac{PM_{tot}^{2}}{2\pi G}\right)^{1/3} \ K\, ,
	\label{eq:Binary_mass_s-star_paper}
\end{equation}

\noindent and inserting the drawn parameters, a distribution of $K$ amplitudes are calculated for this simulated binary star population. When generating a binary system, we also make sure that the system does not result in a merger by calculating the minimum separation and ensuring it does not fall below the radius of the star ($\sim$ 6 R$_{\odot}$). With these simulated distributions for the two populations (see Figure \ref{fig:simulated_k_dist}), we can then use our $K$ amplitude limits - and zero detections - to derive their binary fractions. 



The calculated $K$ amplitude distributions for massive and solar mass stars is for a population made completely of binaries (a binary fraction of 100\%). To make $K$ amplitude distributions for binary fractions less than 100\%, the corresponding percentage of $K$ values are replaced with 0 \kms, representing the single star population. For example, a population with a binary fraction of 50\% will have 50,000 values of 0 \kms, and 50,000 values randomly drawn from the original simulated distribution. $K$ amplitude distributions for different populations with binary fractions ranging from 10 - 100\%, spaced evenly at 10\%, are created. We also conduct finer sampling at binary fractions between 30-50\%.

Once the adjusted $K$ amplitude distribution is established, a simulation is run to determine how many simulated binary star systems would be detected based on our $K$ amplitude limits. For the early-type stars, the $K$ limit from each of our 16 stars in Table \ref{tab:periodicity_results} are compared to a randomly drawn $K$ value from our massive star distribution adjusted for binary fraction. If the drawn $K$ value from the simulated population is higher than the limit from the sample star, we consider it a detection. For each simulation for the massive star population, there can be a minimum of zero detections and a maximum of 16 detections. This simulation is repeated 100,000 times, for each different $K$ amplitude distribution adjusted for binary fraction. The same process is done for the late-type stars using the 12 late-type stars and solar mass $K$ amplitude distributions.

\begin{figure}
\centering
\includegraphics[width=\linewidth]{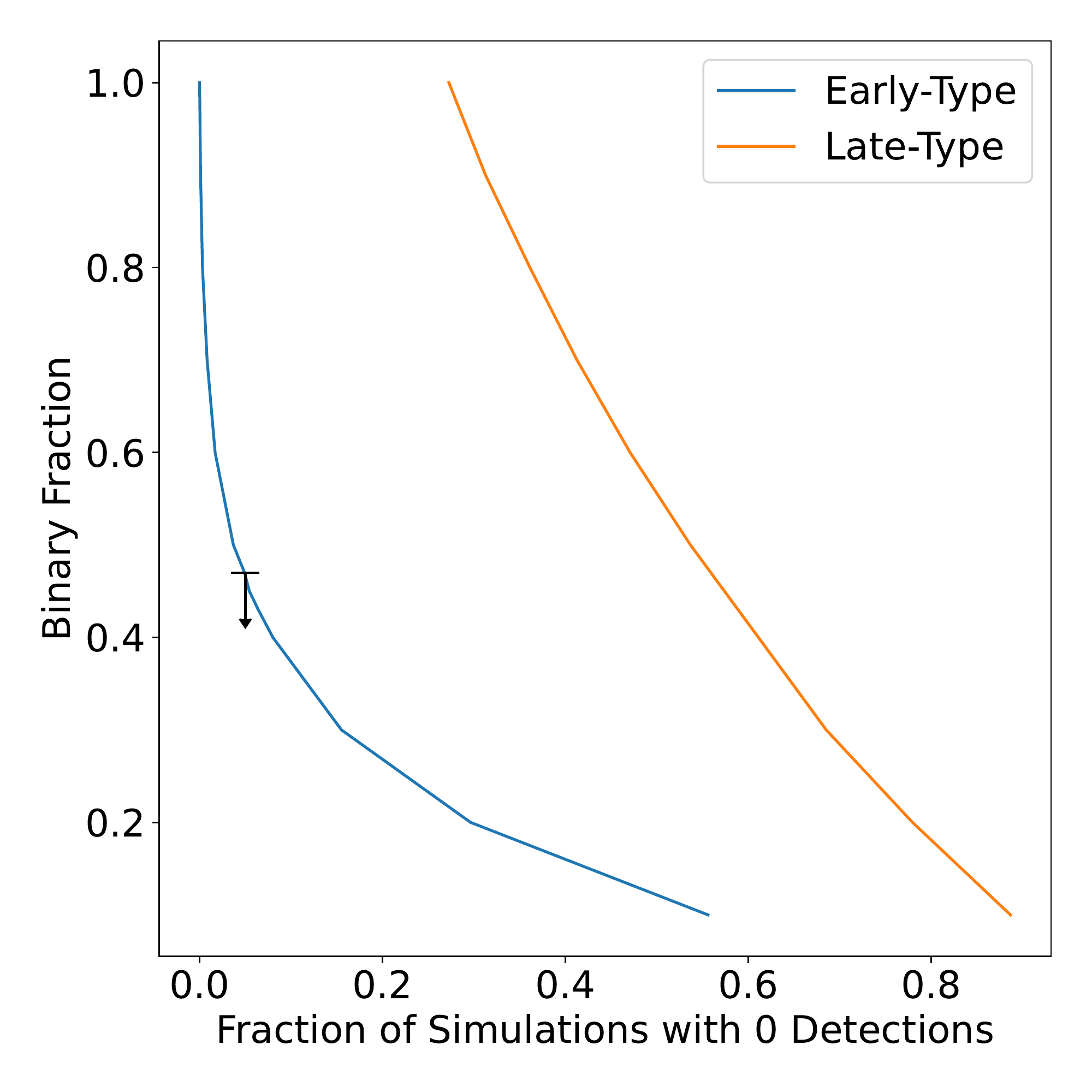}
\caption{
The simulated binary fraction populations versus the fraction of Monte Carlo simulations with zero detections for each population of binary fractions, for both early-type and late-type stars. The binary fraction where we can place an upper limit at 95\% confidence is the black arrow. The early-type star binary fraction limit is 47\%. We cannot place a limit for the late-type stars. Our constraining power lies with the early-type stars, as the $K$ amplitude distribution contains higher values of $K$ compared to the late-type distribution.
\label{fig:binary_compare}
}
\end{figure}

The fraction of simulations with zero detections for each adjusted $K$ amplitude distribution are shown in Figure \ref{fig:binary_compare}. For a massive star population with a 47\% binary fraction, 5\% of the of simulations yielded zero detections. Based on this simulation and our zero binary detections, we can exclude a binary fraction greater than 47\% for this population with a 95\% confidence limit. For the solar mass star populations, a constraint cannot be obtained, with even a 100\% binary fraction only excluded at a 70\% confidence limit.

\section{Discussion} \label{sec:discussion}

\begin{figure}
\centering
\includegraphics[width=\linewidth]{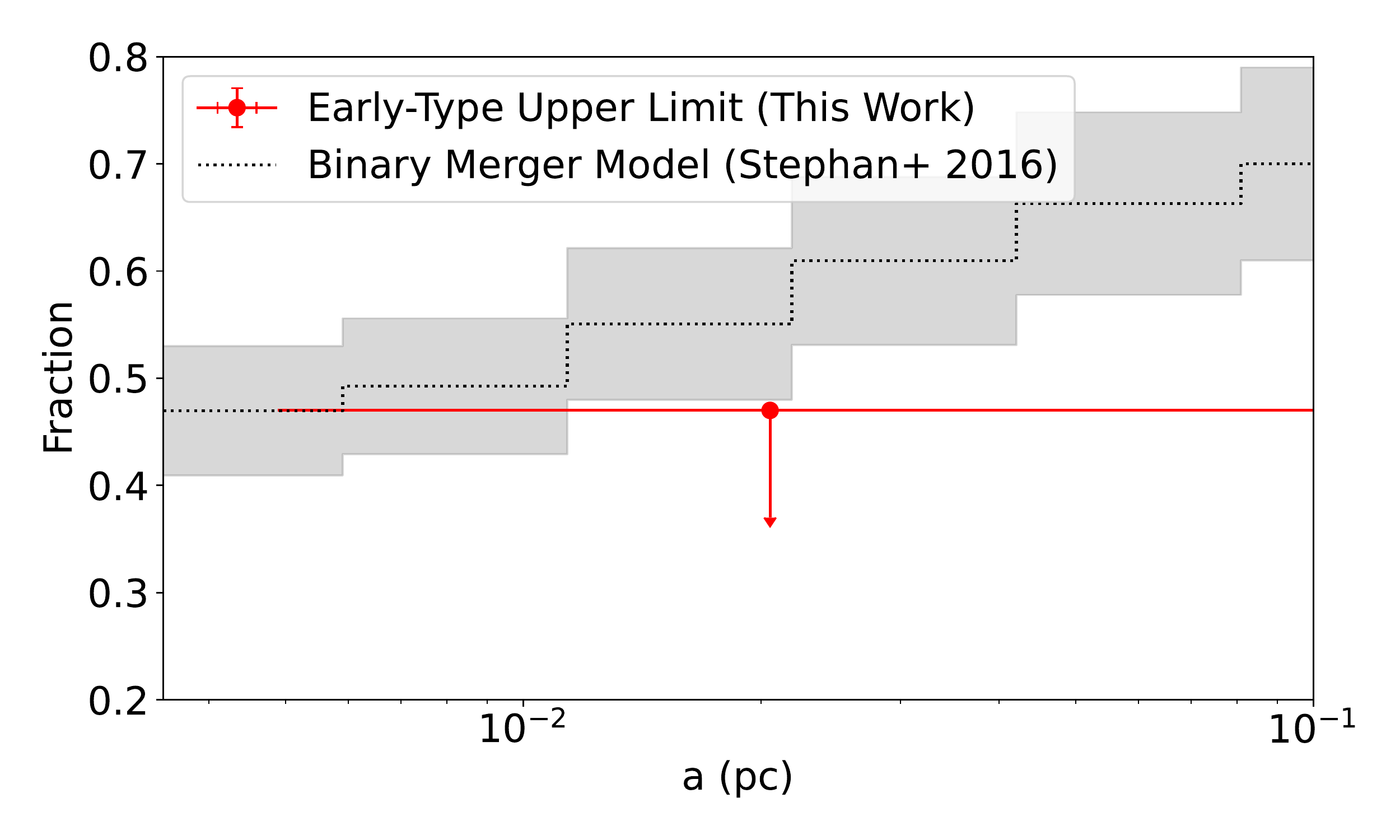}
\caption{
Binary fraction upper limit of 47\% for the early-type star sample (red). The x-axis bar shows the range of the semi-major axis distribution. This limit is compared to the binary fraction model from \citet{Stephan:2016eh} for a given semi-major axis from Sgr A*, normalized to a starting binary fraction limit and uncertainty of 70\%$\pm$9\% for massive stars from \citet{Sana:2012} (gray shaded region). The binary fraction limit from this work for the young stars is consistent with the binary merger model and inconsistent with the binary fraction for massive stars in the solar neighborhood.
\label{fig:stephan_compare}
}
\end{figure}

Our simulations have enabled us to place a limit on the young star population binary fraction at 47\% (with 95\% confidence). This is well below the binary fraction ($70\pm 9\%$) for massive stars larger galactic radii \citep{Sana:2012}. \citet{Stephan:2016eh} have estimated the decrease in the binary star fraction from evaporation and mergers via three-body interactions with the central black hole through the eccentric Kozai-Lidov effect. Figure \ref{fig:stephan_compare} shows their simulation results at an age of 6 Myr, which we normalize to the observed binary star fraction of 70\% at large radii (shaded region). These predictions are consistent with our observations. We also note that the eclipsing binary fraction of stars outside the central arcsecond ($\sim 0.4$ pc) reported by \citet{Gautam:2019} is consistent with the field star binary fraction. The low binary fraction within $\sim$20 mpc appears to be well-explained by a scenario in which the central SMBH drives binary star mergers near its proximity. The process has important implications for the production of gravitational wave sources \citep{LIGO2016_offical}. Additional observations will further improve limits on the multiplicity of these stars closest to the SMBH.

This result of a low binary fraction is also consistent with the binary star disruption mechanism. In this evolution mechanism, a binary star system is tidally disrupted by the SMBH, leaving one single component bound to the SMBH \citep[e.g.][]{Hills:1988br,Perets:2007fo, Fragione2017MNRAS, Generozov2020ApJ}. The other component is ejected as a hypervelocity star, which have been observed in the Milky Way \citep[see][for a review]{Brown:2015}. It is also possible that a triple system may be disrupted by the SMBH and leave behind a captured binary S-star, so the discovery an S-star binary could support a disrupted triple system hypothesis \citep{Fragione2018MNRAS}.

\citet{Naoz2018} explain that unaccounted binary stars can bias the inferred kinematic properties of the nearby clockwise disk of young stars. While the stars in this work are not members of the clockwise disk, it is interesting to compare the young S-stars to the disk population \citep[e.g.][]{Madigan:2014fp}. Given the closer proximity to the SMBH compared to the disk, the S-stars would be more sensitive to the effects of the SMBH. This closer proximity could lead to binary mergers and binary disruptions. Therefore, the S-star binary star fraction can be lower than the disk binary fraction. 

It is not surprising that our binary fraction limit for the late-type stars is not as constraining as the limit for the early-type stars. The late-type stars' $K$ amplitude distribution is dominated by very low values due to the binary population having longer periods and lower stellar masses. Even though we can place lower $K$ amplitude limits for the individual late-type stars given our better radial velocity precision, these lower limits do not outweigh the population's distribution of $K$ amplitudes. Additionally, not identifying binary candidates among the late-type stars is unsurprising. \citet{Stephan:2016eh,Stephan:2019} reports that the evaporation timescale for a binary system with a total mass of 2\msun \ and separated by 3 AU (P $\sim 1300$ days) evaporates in under $10^{6}$ years. Since these late-type stars are $\sim$ 1 Gyr old, these stars have had sufficient time to evaporate, if they were previously part of binary star systems. After a Gyr, \citet{Stephan:2016eh} explains that there has been more time for mergers to take place, so even though binary star systems can survive longer than the evaporation time due to hardening interactions, these hardened, close binary stars can merge as they evolve off the main-sequence\footnote{These merged stars would also appear younger by comparison.}. Nevertheless, discovering binary star systems among the late-type star population would provide a strong constraint for the density of objects at the Galactic center \citep{Rose:2020}, and continued monitoring will provide improved sensitivity for the late-type star population.

\section{Acknowledgements}

 We are grateful for the helpful and constructive comments from the referee. We thank M. R. Morris for his comments and long-term efforts on the Galactic Center Orbits Initiative. The primary data for this work was collected with the W. M. Keck Observatory, which is operated as a scientific partnership among the California Institute of Technology, the University of California, and the National Aeronautics and Space Administration. We wish to recognize that the summit of Maunakea has always held a very significant cultural role for the indigenous Hawaiian community. We are most fortunate to have the opportunity to observe from this mountain. We also thank the staff of the Keck Observatory, especially Jim Lyke, Randy Campbell, Gary Puniwai, Heather Hershey, Hien Tran, Scott Dahm, Jason McIlroy, Joel Hicock, and Terry Stickel, for all their help in obtaining the new observations.  Finally, we are grateful for the financial support for this work provided by NSF AST grants 1412615 and 1909554, the Gordon \& Betty Moore Foundation, the Levine-Leichtman Family Foundation, Ken and Eileen Kaplan Student Support Fund, the Galactic Center Board of Advisors, and the Janet Marott Student Travel Awards. S.N. acknowledges the partial support from NASA ATP 80NSSC20K0505 and thanks Howard and Astrid Preston for their generous support.

\facility{W. M. Keck Observatory, Gemini North Observatory}
\software{Numpy \citep{numpy2011CSE....13b..22V,numpyharris2020array}, Astropy \citep{2013Astropy,Astropy2018AJ}, Starkit \citep{starkit:2015}, gatspy \citep{VanderPlas:2015, VanderPlas:2018}, IRAF \citep{IRAF1986SPIE,IRAF1993ASPC}, Multinest \citep{Feroz:2008fi,Feroz:2009aa,Feroz:2013aa}, SPISEA \citep{spisea2020,Hosek:2020AJ}, Scipy \citep{2020SciPy}, OSIRIS Data Reduction Pipeline \citep{OSIRIS_pipeline:2017, Lockart:2019}}



\appendix




\section{Source Confusion}	\label{sect:confusion_gas_appendix}

We took extra care to ensure that radial velocity measurements were not affected by either stellar or gaseous source confusion. Stellar source confusion affects 23 stars, which are therefore removed from the sample. Local gas can also affect the measurement of the \brg absorption line, since it not only emits \brg, but it does so at different velocities. One of the checks we conducted was to look at the strength of the gas emission at the star's radial velocity in the subtracted background. This led to the removal of two further stars, S1-2 and S1-33, since they were identified as having potentially biased radial velocity measurements based on their subtracted gas backgrounds. Table \ref{tab:excluded_stars} lists the complete list of stars that were excluded from this analysis for all the reasons discussed in Section \ref{sec:Sample}.




\input{excluded_sample}

\section{Impact of OSIRIS detector upgrade}
\label{sect:instrument}

\begin{figure}
\centering
\includegraphics[width=\linewidth]{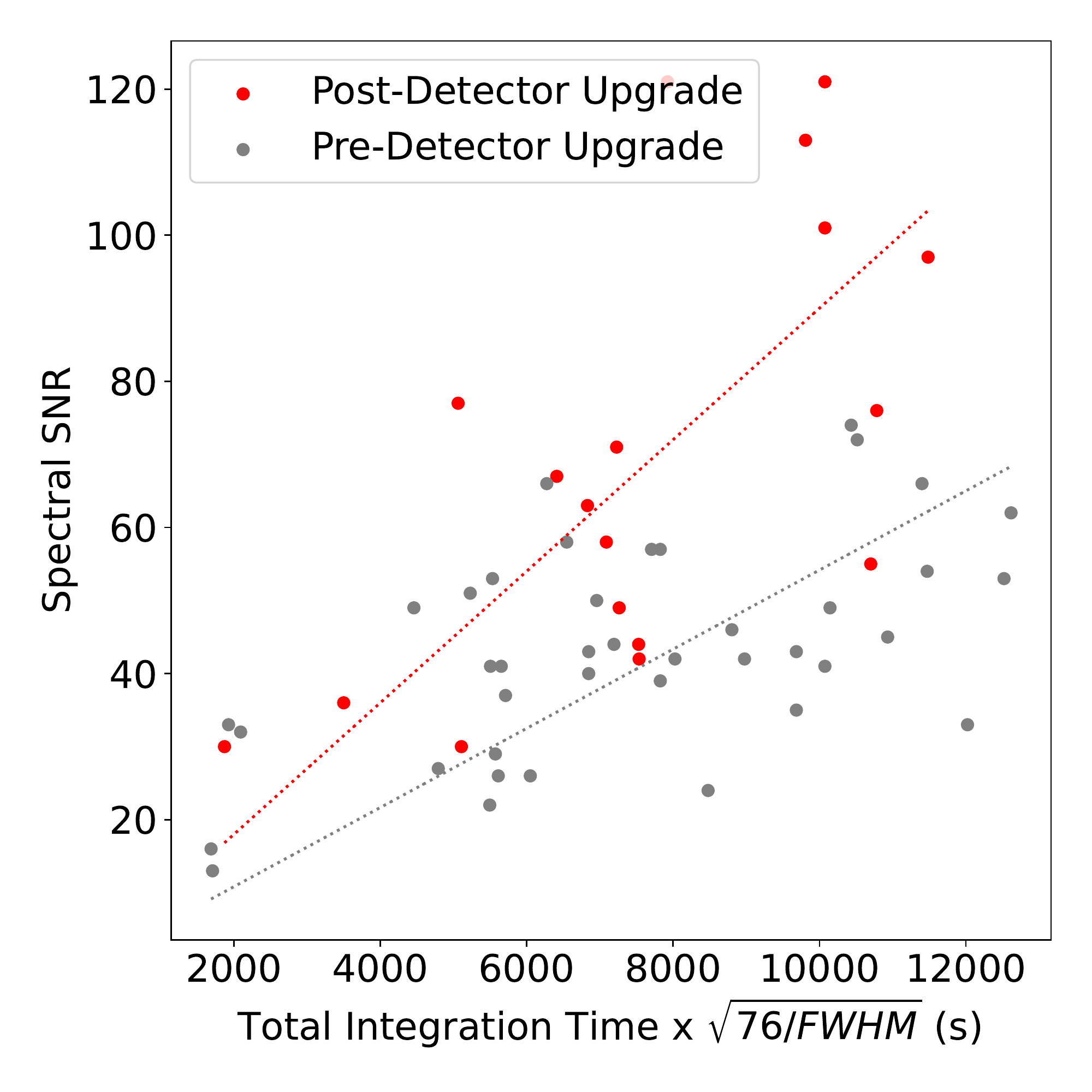}
\caption{
Spectral signal-to-noise ratio of a $K'$ $\sim$14 mag star for a dataset's total integration time scaled by the dataset's FWHM relative to the average FWHM of 76 mas. Data taken with the newest OSIRIS detector and previous detector are plotted in red and gray, respectively. The dashed lines are fits to the data subsets. The steeper slope of the new detector data fit ($9.00\pm 0.69\times10^{-3}$) versus the old detector data ($5.42 \pm 0.29\times10^{-3}$) shows the improved spectral signal-to-noise for a given integration time and FWHM.
\label{fig:tint_fwhm}
}
\end{figure}

Figure \ref{fig:tint_fwhm} compares the performance of the old and new detector for a K $\sim$14 star from our standard Galactic Center observational set-up. The new detector has enabled improved spectral signal-to-noise for data for a given total integration time and FWHM.

\section{Lomb-Scargle and $K$ Amplitude Limits}
\label{sect:ls_k_results}

\begin{figure*}
\centering
\includegraphics[width=\linewidth]{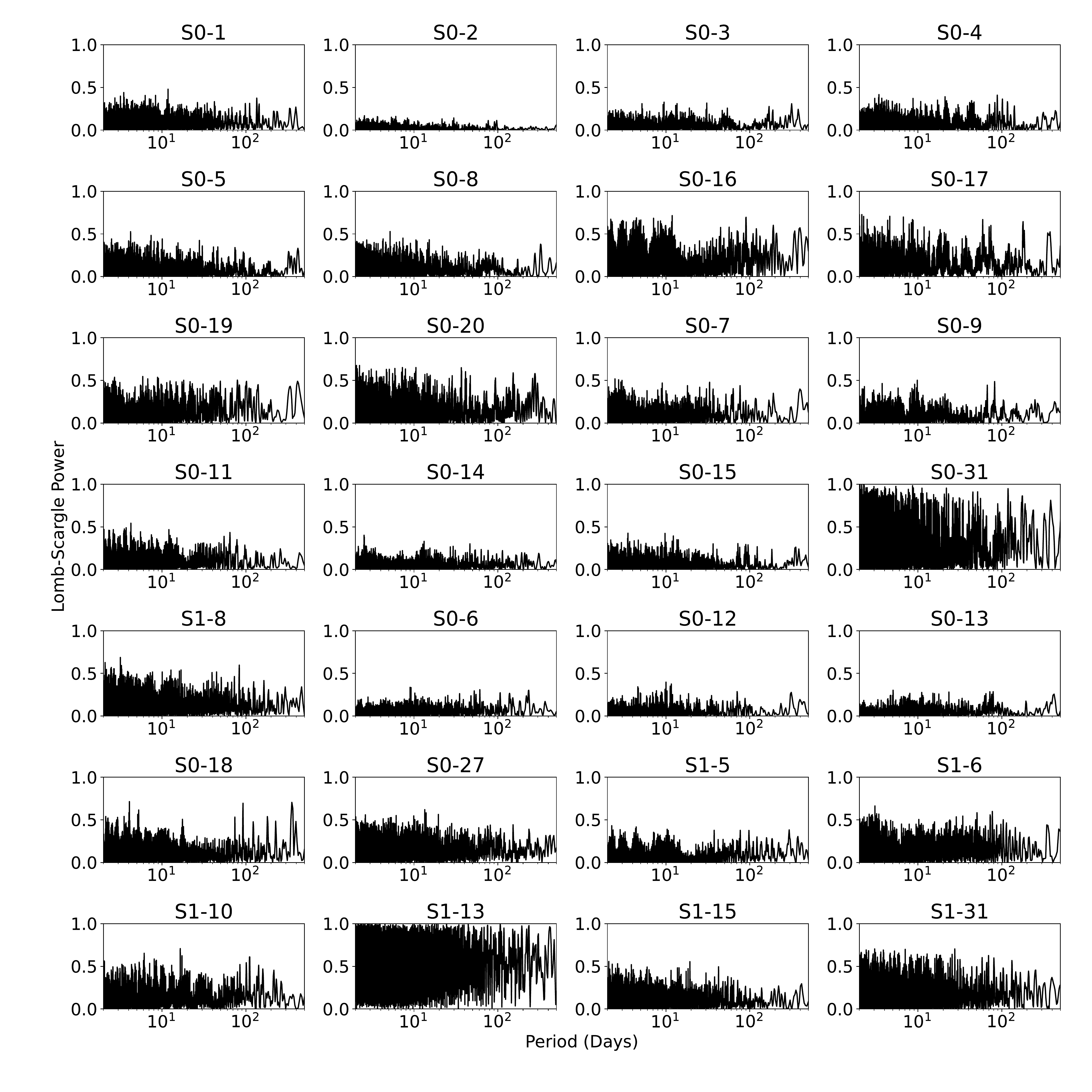}
\caption{
Lomb-Scargle periodograms for all 28 stars in the sample. For each plot, period in days is plotted on the x-axis, and the Lomb-Scargle power is on the y-axis.
\label{fig:all_ls}
}
\end{figure*}

\begin{figure*}
\centering
\includegraphics[width=\linewidth]{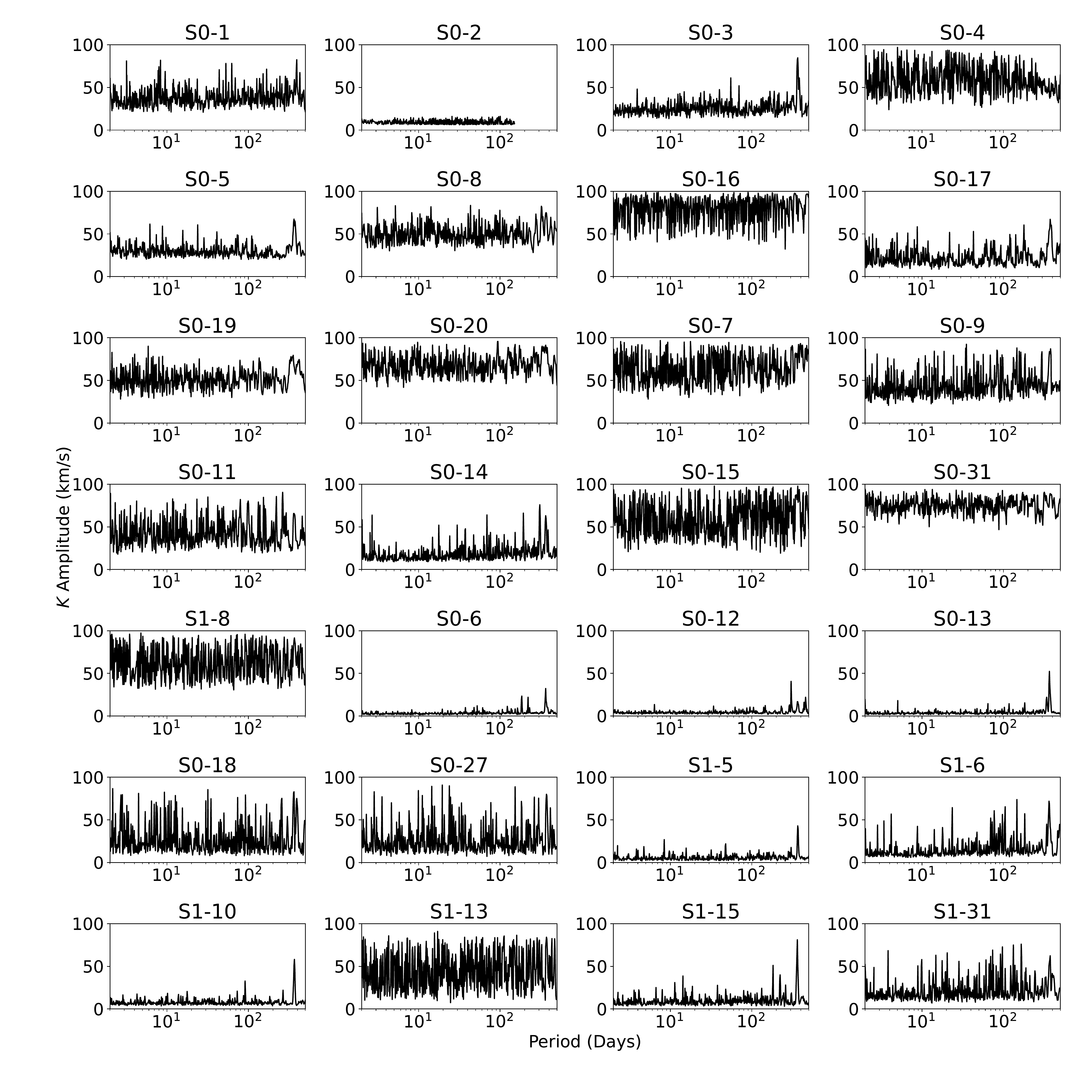}
\caption{
$K$ amplitude limits for all 28 stars in the sample. For each plot, period in days is plotted on the x-axis, and the $K$ amplitude limit in \kms is on the y-axis.
\label{fig:all_k}
}
\end{figure*}

This appendix section presents the results discussed in Section \ref{sec:per_search}. Figure \ref{fig:all_ls} shows the Lomb-Scargle periodograms for all stars used in the analysis, and Figure \ref{fig:all_k} shows the $K$ amplitude limits per period.

\section{Placing limits on companion masses}
\label{sec:comp_mass_limits}

\begin{figure}
\centering
\includegraphics[width=\linewidth]{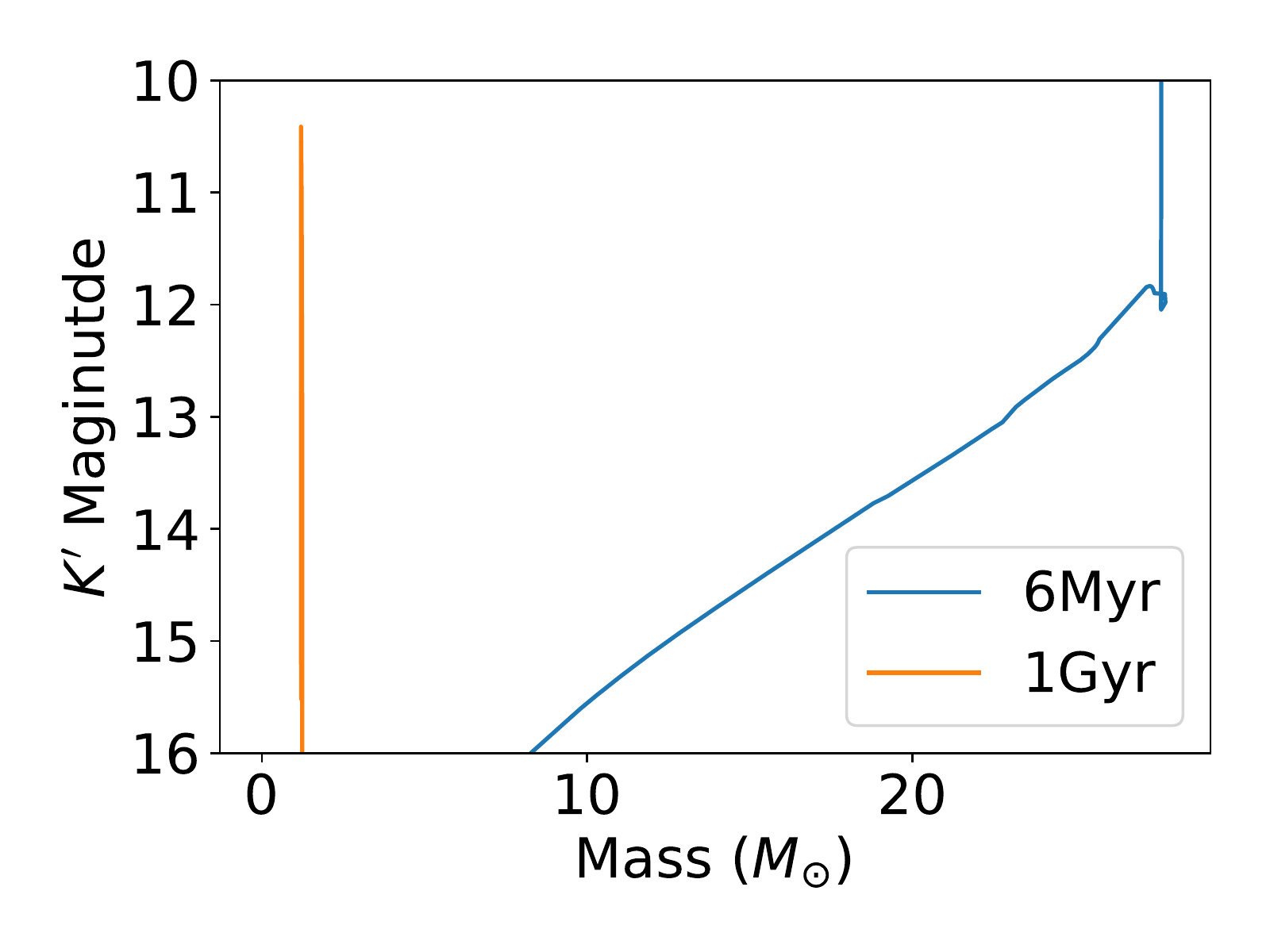}
\caption{
Two SPISEA isochronese used for determining the mass of a star based on its $K'$ magnitude. The MIST stellar evolution models \citep{MIST2016ApJ} and extinction law from \citep{schodel:2010} are applied to these isochrones. The 6Myr isochrone was used for the early-type stars, while the 1 Gyr isochrone was used for the late-type stars.
\label{fig:isochrones}
}
\end{figure}

\begin{figure}
\centering
\includegraphics[width=\linewidth]{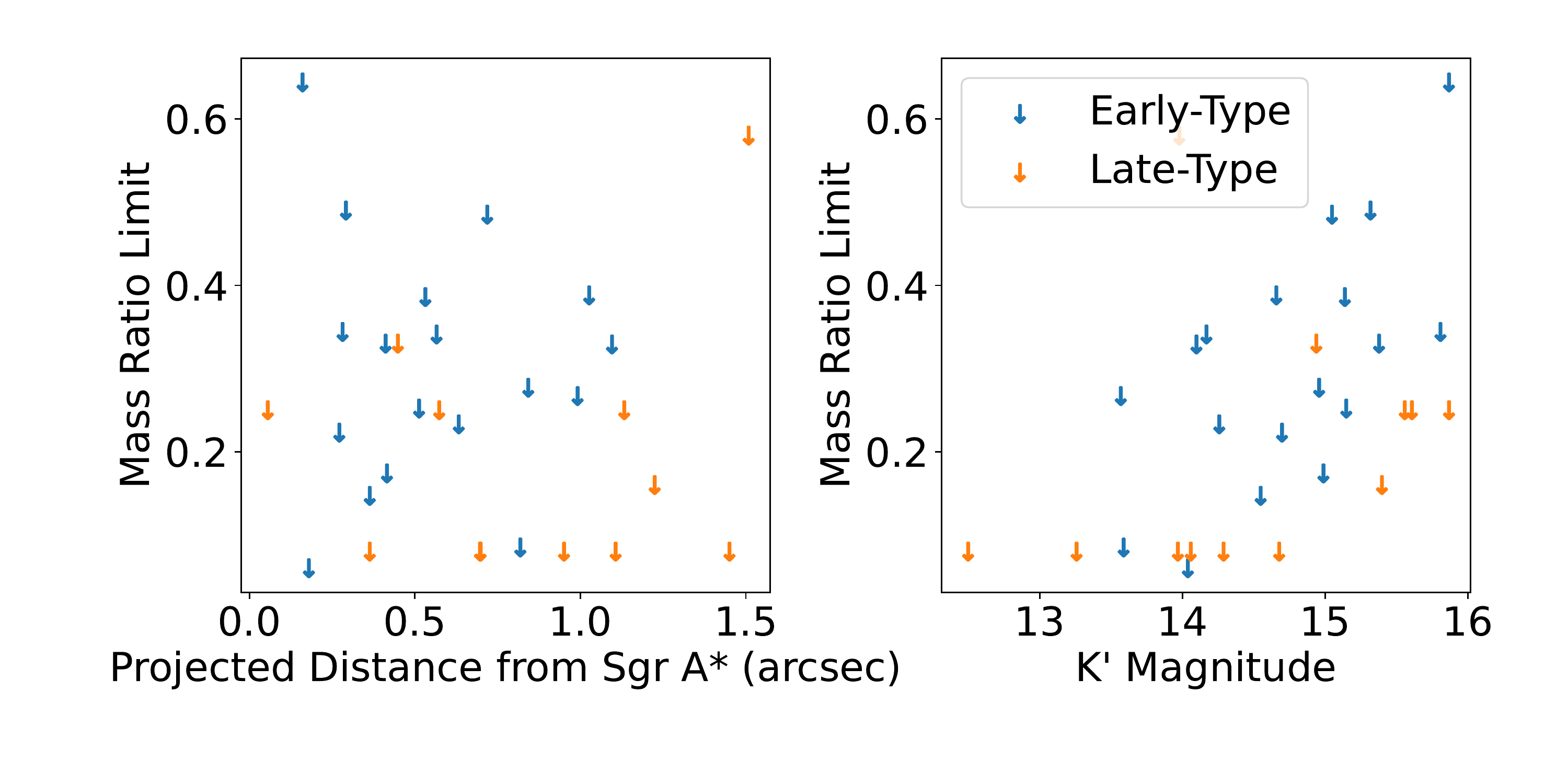}
\caption{
Left: Companion mass limits for each star plotted with their projected distance from Sgr A*. These limits come from marginalizing the mass limits over all sampled periods. Right: Companion mass limits for each star plotted with their $K'$ magnitude.
\label{fig:Kp_mass}
}
\end{figure}

With the results from the binary curve fitting, in particular our limits on the $K$ amplitude, we move to place limits on hypothetical companion masses of binary systems using the same methodology as \citet{Chu:2018}. For each period $P$, there is a limit on $K$, and the binary mass equation (Equation \ref{eq:Binary_mass_s-star_paper} can be solved assuming for a total mass, a limit for the companion mass for each period can be calculated. In order to determine the total mass for a star, we use its $K'$ photometry reported \citet{Gautam:2019} and an isochrone generated with the SPISEA software \citep{spisea2020,Hosek:2020AJ}. A 6.78 Myr isochrone is used for the early-type stars and a 1 Gyr isochrone is used for the late-type stars. These isochrones use the MIST stellar evolution models \citep{MIST2016ApJ}, and each isochrone is corrected for extinction to the Galactic Center with a value of $A_{K'} = 2.46$ \citep{schodel:2010}. Solar metalicities are used for both isochrones. These isochrones are shown in Figure \ref{fig:isochrones}. The total mass used for each star is given in Table \ref{tab:mass_limits}. \citet{Habibi:2017} reported masses for early-type S-stars stars in their analysis. For stars that overlap with our sample, their reported mass values are lower than the isochrone mass values but still consistent within 2$\sigma$. We report the median upper limits for the companion masses for all periods in Table \ref{tab:mass_limits} and Figure \ref{fig:Kp_mass}.


\input{mass_results_appendix}

The photometric information from \citet{Gautam:2019} of each star is used to place conservative limits on the masses of an equal mass binary system. To do this, the total flux from the star is divided in half. The SPISEA isochrone is then searched to find the mass of a star that would contribute the equivalent amount of flux. This places a limit on the components of a face-on binary system composed of equal mass stars. This can be thought of as a conservative limit, as the spectral differences between different mass stars are not considered in this part of the analysis. These limits are reported in Table \ref{tab:mass_limits}.




\section{Effect of Period Distributions and Sample Size on Simulations}
\label{sec:period_simulations}

\begin{figure}
\centering
\includegraphics[width=\linewidth]{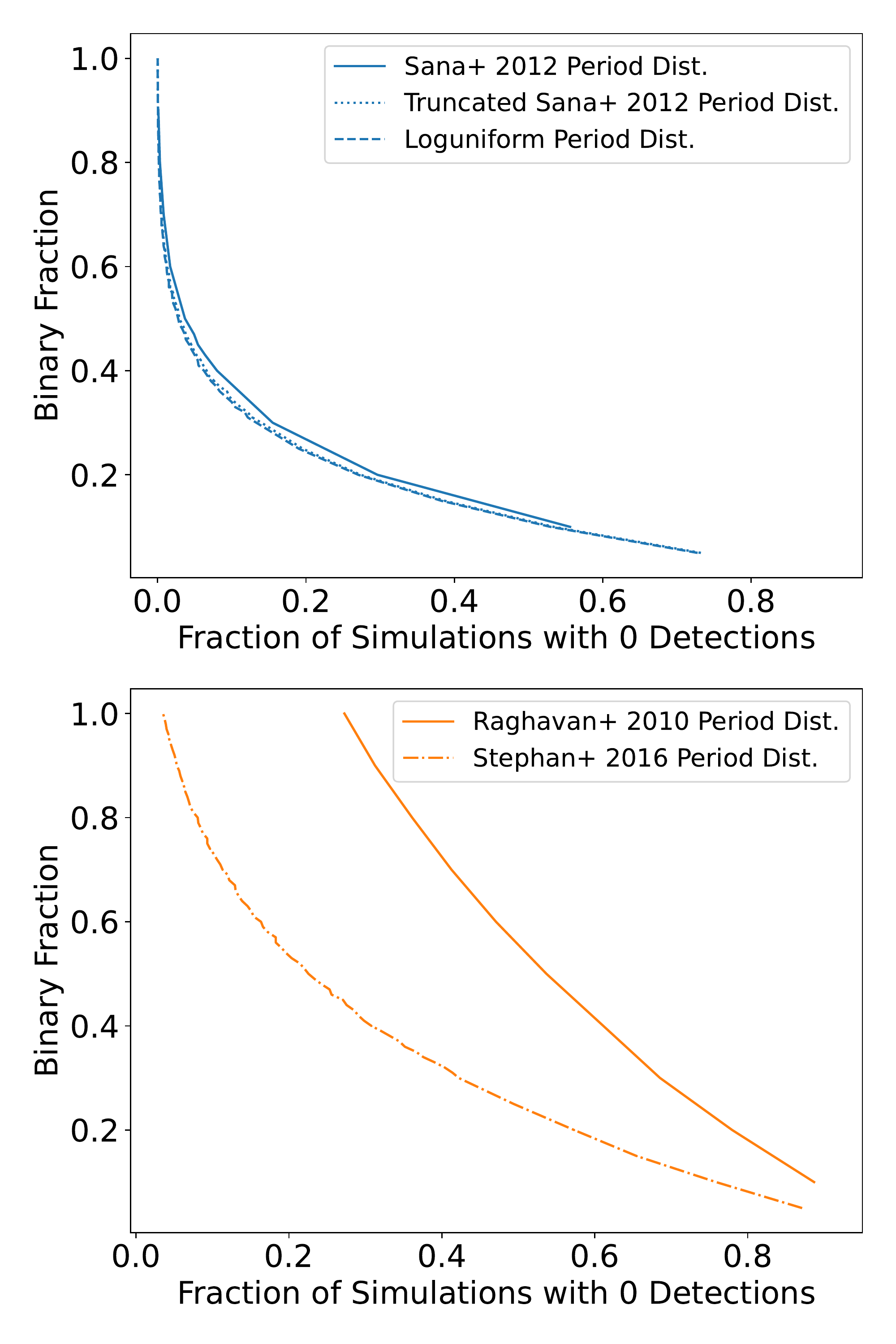}
\caption{
The simulated binary fraction populations versus the fraction of Monte Carlo simulations with zero detections for each population of binary fractions with different period distributions. Left: The early-type star simulations when using the following period distributions: \citet{Sana:2012}, truncated \citet{Sana:2012}, and loguniform. The binary fraction limit changes by less than 3\%. Right: The late-type star simulations when using the following period distributions: \citet{Raghavan:2010} and \citet{Stephan:2016eh}. The \citet{Stephan:2016eh} distribution truncates the longer period binaries compared to \citet{Raghavan:2010}. We would infer a binary fraction less than 93\% at 95\% confidence with the \citet{Stephan:2016eh} distribution.
\label{fig:period_investigation}
}
\end{figure}

We note that while we adopt field-like distributions for the binaries, there are processes that truncate long period binaries in the dense environment of the Galactic Center compared to the field. For example, flyby stars unbind wide separated binaries \citep[e.g.][see latter for unbinding of eccentric binaries]{BinneyTremaine2008, Rose:2020} which produces a distribution described by \citet{Stephan:2016eh,Stephan:2019}. Furthermore, stability and the Hills process tend to truncate the distribution in a similar manner. Thus, it can be estimated as loguniform in the literature \citep[e.g.][]{Fragione2019MNRAS}. Note that the \citet{Sana:2012} distribution, adopted here for the early-type stars, favors short-period distribution, consistent with the unbinding processes \citep[as noted in][]{Hoang2018}. We examine the impact of these effects on our results by testing two period distributions for the young stars: (1) the \citet{Sana:2012} period distribution truncated at $9.8 \times 10^{4}$ days, (2) a lognormal distribution truncated at the same length. We find that both of these period distributions produce no impact on our inference about the binary fraction. The binary fraction limit changes by less than 3\%. The effects of these period distributions are shown in Figure \ref{fig:period_investigation}.

For late-type stars, the binary distribution makes a modest difference in the resulting binary fraction limit because the \citet{Raghavan:2010} distribution extends to very long periods that these dynamical processes will truncate. To approximate these effects, we use the results of the simulations from \citet{Stephan:2016eh} which shows a truncated period distribution at $3.3 \times 10^{5}$ days for 1.2 \msun stars. If the late-type star binary population is truncated, then our data has a modest constraint on the binary fraction. We would infer an upper limit of the binary fraction to be less than 93\% at 95\% confidence. 

We have also explored our sensitivity to sample size for our simulations. To account for an increased sample size, we double-count stars from our observational sample and compare them to the simulated populations. For the young stars, it would take an increase of eight stars (from 16 to 24) to decrease the binary fraction limit by 10\%. Changing the sample by one or two stars for the early-type stars does not dramatically affect the limit.

\bibliography{apj-jour,sstar_paper.bib}


\end{CJK*}
\end{document}

%% file: all_spec_table_v2_09-28-22.tex
\startlongtable
\begin{deluxetable*}{lccccccccc}
\tablecolumns{10} 
\tablewidth{0pc} 
\tablecaption{Summary of Spectroscopic Observations with New Radial Velocities\label{tab:all_spec_obs}}
\tablehead{ 
    \multicolumn{3}{c}{Date\tna}  & 
    \colhead{Instrument} &
	\colhead{$N_{\text{frames}} \times t_{\text{int}}$} &
	\colhead{Filter} &
	\colhead{Scale} &
	\colhead{FWHM\tnb} &
	\colhead{SNR\tnb} &
	\colhead{New RVs} \\
	\cline{1-3}
    \colhead{(UT)} &
	\colhead{(MJD)} &
    \colhead{(Epoch)} &
	\colhead{} &
	\colhead{(s)} &
	\colhead{} &
	\colhead{(mas)} &
	\colhead{(mas)} &
	\colhead{} &
	\colhead{This Work\tnc} 
}
\startdata
2005-07-03$^{1}$		&  	53554.50	&  	2005.503	&	OSIRIS	&  	7  $\times$ 900   	&	Kbb 	&	20	&	58	  	&	44	&	5\\
2006-06-18$^{1}$		&  	53904.50	&  	2006.461	&	OSIRIS	&  	9  $\times$ 900   	&	Kn3 	&	35	&	81	  	&	39	&	10\\
2006-06-30$^{1}$		&  	53916.50	&  	2006.494	&	OSIRIS	&  	9  $\times$ 900   	&	Kn3 	&	35	&	77	  	&	42	&	12\\
2006-07-01$^{1}$		&  	53917.50	&  	2006.497	&	OSIRIS	&  	9  $\times$ 900   	&	Kn3 	&	35	&	64	  	&	46	&	8\\
2007-05-21$^{1}$		&  	54241.50	&  	2007.384	&	OSIRIS	&  	2 $\times$ 900   	&	Kn3 	&	35	&	86		&	16	&	5\\
2007-07-18$^{1}$		&  	54299.29	&  	2007.542	&	OSIRIS	&  	2 $\times$ 900   	&	Kn3 	&	35	&	66$^{5}$		&	33$^{5}$	&	4\\
2007-07-19$^{1}$		&  	54300.29	&  	2007.545	&	OSIRIS	&  	2 $\times$ 900   	&	Kn3 	&	35	&	56	  	&	32	&	6\\
2008-05-16$^{2}$		&  	54602.50	&  	2008.372	&	OSIRIS	&  	11 $\times$ 900   	&	Kn3 	&	35	&	57	  	&	66	&	17\\
2008-07-25$^{2}$		&  	54672.28	&  	2008.563	&	OSIRIS	&  	9  $\times$ 900 	&	Kn3 	&	35	&	81	  	&	57	&	17\\
2009-05-05$^{2}$		&  	54956.50	&  	2009.342	&	OSIRIS	&  	7   $\times$ 900	&	Kn3 	&	35	&	70	  	&	58	&	18\\
2009-05-06$^{2}$		&  	54957.50	&  	2009.344	&	OSIRIS	&  	12  $\times$ 900	&	Kn3 	&	35	&	81	  	&	74	&	9\\
2010-05-05$^{2}$		&  	55321.50	&  	2010.341	&	OSIRIS	&  	6  $\times$ 900 	&	Kn3 	&	35	&	70	  	&	26	&	8\\
2010-05-08$^{2}$		&  	55324.50	&  	2010.349	&	OSIRIS	&  	11  $\times$ 900 	&	Kn3 	&	35	&	79	  	&	43	&	19\\
2011-07-10$^{2}$		&  	55752.33	&  	2011.520	&	OSIRIS	&  	6   $\times$ 900   	&	Kn3 	&	35	&	71	  	&	29	&	20\\
2011-07-19$^{2}$		&  	55761.31	&  	2011.545	&	OSIRIS	&  	6 $\times$ 900   	&	Kn3 	&	35	&	96$^{6}$		&	27$^{6}$	&	7\\
2012-06-11$^{2}$		&  	56089.50	&  	2012.444	&	OSIRIS	&  	7  	$\times$ 900	&	Kn3 	&	20	&	64	  	&	40	&	4\\
2012-07-22$^{2}$		&  	56130.31	&  	2012.555	&	OSIRIS	&  	7   $\times$ 900   	&	Kn3 	&	35	&	92	 	&	37	&	10\\
2012-08-12$^{2}$		&  	56151.33	&  	2012.613	&	OSIRIS	&  	6   $\times$ 900   	&	Kn3 	&	35	&	56	  	&	66	&	6\\
2012-08-13$^{2}$		&  	56152.27	&  	2012.615	&	OSIRIS	&  	7   $\times$ 900	&	Kn3 	&	35	&	99	  	&	41	&	8\\
2013-05-11$^{2}$		&  	56423.50	&  	2013.358	&	OSIRIS	&  	11  $\times$ 900	&	Kbb		&	35	&	73	  	&	41	&	7\\
2013-05-12$^{2}$		&  	56424.50	&  	2013.361	&	OSIRIS	&  	11  $\times$ 900  	&	Kbb 	&	35	&	62	  	&	45	&	3\\
2013-05-13$^{2}$		&  	56425.50	&  	2013.363	&	OSIRIS	&  	12  $\times$ 900	&	Kbb	 	&	35	&	61	  	&	33	&	3\\
2013-05-14$^{2}$		&  	56426.50	&  	2013.366	&	OSIRIS	&  	11  $\times$ 900   	&	Kn3	 	&	35	&	67	  	&	72	&	21\\
2013-05-16$^{2}$		&  	56428.50	&  	2013.372	&	OSIRIS	&  	7  $\times$ 900		&	Kn3	 	&	20	&	98	  	&	53	&	5\\
2013-05-17$^{2}$		&  	56429.50	&  	2013.374	&	OSIRIS	&  	7  $\times$ 900 	&	Kn3	 	&	20	&	64   	&	43	&	5\\
2013-07-25$^{2}$		&  	56498.33	&  	2013.563	&	OSIRIS	&  	11  $\times$ 900 	&	Kn3	 	&	35	&	79   	&	35	&	8\\
2013-07-26$^{2}$		&  	56499.34	&  	2013.566	&	OSIRIS	&  	6  $\times$ 900 	&	Kn3	 	&	35	&	73   	&	22	&	5\\
2013-07-27$^{2}$		&  	56500.33	&  	2013.568	&	OSIRIS	&  	11  $\times$ 900 	&	Kn3	 	&	35	&	72   	&	49	&	18\\
2013-08-10$^{2}$		&  	56514.29	&  	2013.607	&	OSIRIS	&  	7 $\times$ 900		&	Kn3 	&	35	&	62   	&	50	&	13\\
2013-08-11$^{2}$		&  	56515.31	&  	2013.609	&	OSIRIS	&  	9 $\times$ 900		&	Kn3 	&	35	&	69   	&	24	&	14\\
2013-08-13$^{2}$		&  	56517.29	&  	2013.615	&	OSIRIS	&  	12 $\times$ 900		&	Kn3 	&	35	&	67   	&	54	&	11\\
2014-05-17				&  	56794.51	&  	2014.374	&	OSIRIS	&  	6 $\times$ 900   	&	Kn3 	&	35	&	69$^{6}$		&	41$^{6}$	&	6\\
2014-05-18$^{3}$		&	56795.50	& 	2014.376	&	OSIRIS	& 	13 $\times$ 900		&	Kn3		&	35	& 	66 		&	53	&	18\\
2014-05-19				&  	56796.51	&  	2014.379	&	OSIRIS	&  	13 $\times$ 900   	&	Kn3 	&	35	&	65$^{6}$		&	62$^{6}$	&	20\\
2014-05-22				&  	56799.51	&  	2014.387	&	OSIRIS	&  	7 $\times$ 900   	&	Kn3 	&	35	&	82$^{6}$		&	26$^{6}$	&	6\\
2014-05-23$^{3}$		&	56800.50	& 	2014.390	&	OSIRIS	& 	10 $\times$ 900		&	Kn3		&	35	& 	76 		&	42	&	8\\
2014-07-03$^{3}$		&	56841.36	& 	2014.502	&	OSIRIS	& 	8 $\times$ 900		&	Kn3		&	35	& 	66 		&	57	&	22\\
2015-05-04$^{3}$		&	57146.50	& 	2015.337	&	OSIRIS	& 	5 $\times$ 900		&	Kn3		&	35	& 	77 		&	49	&	4\\
2015-07-21$^{3}$		&	57224.35	& 	2015.551	&	OSIRIS	& 	5 $\times$ 900		&	Kn3		&	35	& 	56 		&	51	&	24\\
2015-08-07				&	57241.33	& 	2015.597	&	OSIRIS	& 	2 $\times$ 900		&	Kn3		&	35	& 	84$^{7}$ 		&	13$^{7}$	&	1\\
2016-05-14$^{3}$		&	57522.50	& 	2016.367	&	OSIRIS	& 	8 $\times$ 900		&	Kbb		&	35	& 	78 		&	58	&	1\\
2016-05-15$^{3}$		&	57523.50	& 	2016.370	&	OSIRIS	& 	4 $\times$ 900		&	Kbb		&	35	& 	80 		&	36	&	2\\
2016-05-16$^{3}$		&	57524.50	& 	2016.372	&	OSIRIS	& 	8 $\times$ 900		&	Kbb		&	35	& 	84 		&	63	&	2\\
2016-07-11				&	57580.35	& 	2016.525	&	OSIRIS	& 	8 $\times$ 900		&	Kbb		&	35	& 	69$^{8}$		&	42$^{8}$ &	6\\
2016-07-12				&	57581.33	& 	2016.528	&	OSIRIS	& 	7 $\times$ 900		&	Kbb		&	35	& 	115$^{9}$		&	30$^{9}$ &	1\\
2017-05-17$^{4}$		&	57890.52	&	2017.374	&	OSIRIS	&	11 $\times$ 900		&	Kn3		&	35 	&	73		&	101	&	20\\
2017-05-18$^{4}$		&	57891.51	&	2017.377	&	OSIRIS	&	9 $\times$ 900		&	Kn3		&	35 	&	94		&	49	&	20\\
2017-05-19$^{4}$		&	57892.50	&	2017.379	&	OSIRIS	&	6 $\times$ 900		&	Kn3		&	35 	&	86		&	77	&	8\\
2017-07-19$^{4}$		&	57953.33	&	2017.546	&	OSIRIS	&	12 $\times$ 900		&	Kn3		&	35 	&	77		&	55	&	10\\
2017-07-27$^{4}$		&	57961.32	&	2017.568	&	OSIRIS	&	13 $\times$ 900		&	Kn3		&	35 	&	89		&	76	&	20\\
2017-08-14$^{4}$		&	57979.28	&	2017.617	&	OSIRIS	&	8 $\times$ 900		&	Kn3		&	35 	&	75		&	71	&	21\\
2018-03-17$^{4}$		&	58194.64	&	2018.207	&	OSIRIS	&	2 $\times$ 900		&	Kn3		&	35 	&	70		&	30	&	6\\
2018-04-24$^{4}$		&	58232.57	&	2018.310	&	OSIRIS	&	7 $\times$ 900		&	Kn3		&	35 	&	73		&	67	&	12\\
2018-05-13$^{4}$		&	58251.51	&	2018.362	&	NIFS	&	12 $\times$ 600		&	K		&	50 $\times$ 100 	&		&	84	&	7\\
2018-05-22$^{4}$		&	58260.49	&	2018.387	&	NIFS	&	7 $\times$ 600		&	K		&	50 $\times$ 100 	&		&	66	&	4\\
2018-05-23$^{4}$		&	58261.50	&	2018.390	&	OSIRIS	&	14 $\times$ 900		&	Kn3		&	35 	&	91		&	97	&	19\\
2018-06-05$^{4}$		&	58274.47	&	2018.425	&	OSIRIS	&	10 $\times$ 900		&	Kn3		&	35 	&	108		&	44	&	4\\
2018-07-22$^{4}$		&	58321.33	&	2018.554	&	OSIRIS	&	11 $\times$ 900		&	Kn3		&	35 	&	77		&	113	&	18\\
2018-07-31$^{4}$		&	58330.32	&	2018.578	&	OSIRIS	&	11 $\times$ 900		&	Kn3		&	35 	&	73		&	121	&	16\\
2018-08-11$^{4}$		&	58341.31	&	2018.608	&	OSIRIS	&	9 $\times$ 900		&	Kn3		&	35 	&	79		&	121	&	19\\
\enddata
\tablenotetext{a}{These observations, where noted, were first reported for studies of S0-2 alone in the following references: 1) \citet{Ghez:2008ty}, 2) \citet{Boehle:2016ko}, 3) \citet{Chu:2018}, 4) \citet{Do:2019}.}
\tablenotetext{b}{The reported values are assessed on S0-2 unless otherwise noted as follows: 5) S1-15, 6) S0-14, 7) S1-13, 8) S0-3, 9) S0-12. All stars used for characterization have $K'$ mag of 13.5-14.5.}
\tablenotetext{c}{This includes only RVs for the final sample.}
\tablecomments{Col 1-3: date of observation given in UT, modified Julian date, and Julian Year, Col 4: instrument name, Col 5: number of frames combined times exposure time of each frame, Col 6: instrument filter, Col 7: pixel scale used, Col 8: FWHM of reference star, Col 9: spectral signal-to-noise ratio of reference star, Col 10: new radial velocity measurements reported.}
\end{deluxetable*}

%% file: full_sample_03-13-23.tex
\begin{deluxetable*}{llcrrrrrrcrr}
\tablecolumns{11}
\tablewidth{0pc} 
\tablecaption{S-star RV Sample\label{tab:full_rv_sample}}
\tablehead{
\colhead{Star}	&
\colhead{$K'$}	&
\colhead{Spectral}	&
\colhead{RA$\Delta$\tna}	&
\colhead{Dec$\Delta$\tna}	&
\colhead{R2D\tna}	&
\colhead{RV}	&
\colhead{[RV $\sigma$]}	&
\colhead{RV Baseline}	&
\colhead{RV Long-term Trend}	&
\colhead{Semimajor Axis}		\\
\colhead{}	&
\colhead{(mag)}	&
\colhead{Type}	&
\colhead{(\arcsec)}	&
\colhead{(\arcsec)}	&
\colhead{(\arcsec)}	&
\colhead{Points}	&
\colhead{(\kms)}	&
\colhead{(Years)}	&
\colhead{Method}	&
\colhead{(mpc)}
}
\startdata
S0-1	&	14.7	&	Early	&	0.04	&	-0.26	&	0.264   &	50	&	53	&	15	&	   Orbit	&	24.43 $\pm$ 0.46   \\
S0-2	&	14.0	&	Early	&	-0.01	&	0.17	&	0.172   &	115	&	23	&	18	&	   Orbit	&	4.885 $\pm$ 0.024    \\
S0-3	&	14.5	&	Early	&	0.34	&	0.12	&	0.356   &	59	&	37	&	14	&	   Orbit	&	14.082 $\pm$ 0.082    \\
S0-4	&	14.1	&	Early	&	0.45	&	-0.33	&	0.558   &	52	&	46	&	15	&	   Orbit	&	16.39 $\pm$ 0.66    \\
S0-5	&	15.0	&	Early	&	0.17	&	-0.36	&	0.408   &	42	&	61	&	14	&	   Orbit	&	10.678 $\pm$ 0.067    \\
S0-7	&	15.1	&	Early	&	0.51	&	0.10	&	0.524   &	23	&	30	&	12	&   Polynomial	&	39.9 $\pm$ 8.0     \\
S0-8	&	15.8	&	Early	&	-0.23	&	0.16	&	0.274   &	45	&	95	&	14	&	   Orbit	&	16.612 $\pm$ 0.089    \\
S0-9	&	14.2	&	Early	&	0.22	&	-0.60	&	0.625   &	33	&	36	&	13	&   Polynomial	&	69 $\pm$ 14     \\
S0-11	&	15.1	&	Early	&	0.49	&	-0.06	&	0.505   &	28	&	32	&	12	&   Polynomial	&	103 $\pm$ 21     \\
S0-14	&	13.5	&	Early	&	-0.76	&	-0.28	&	0.811   &	41	&	18	&	12	&   Polynomial	&	48.7 $\pm$ 9.7     \\
S0-15	&	13.5	&	Early	&	-0.97	&	0.18	&	0.984   &	31	&	32	&	12	&   Polynomial	&	54 $\pm$ 11     \\
S0-16	&	15.3	&	Early	&	0.23	&	0.17	&	0.284   &	24	&	76	&	14	&	   Orbit	&	11.611 $\pm$ 0.062    \\
S0-19	&	15.3	&	Early	&	-0.08	&	0.40	&	0.404   &	39	&	120	&	15	&	   Orbit	&	11.581 $\pm$ 0.040    \\
S0-20	&	15.8	&	Early	&	0.05	&	0.14	&	0.153   &	32	&	200	&	14	&	   Orbit	&	10.260 $\pm$ 0.033    \\
S0-31	&	14.9	&	Early	&	0.57	&	0.45	&	0.711   &	9	&	41	&	11	&   Polynomial	&	57 $\pm$ 11     \\
S1-8	&	14.0	&	Early	&	-0.58	&	-0.92	&	1.088   &	16	&	33	&	11	&	   Polynomial	&	89 $\pm$ 18     \\
\hline
S0-6	&	14.0	&	Late	&	0.02	&	-0.36	&	0.356   &	47	&	2.9	&	13	&   Polynomial	&	102 $\pm$ 20     \\
S0-12	&	14.3	&	Late	&	-0.55	&	0.41	&	0.689   &	48	&	3.1	&	12	&   Polynomial	&	115 $\pm$ 23     \\
S0-13	&	13.2	&	Late	&	0.56	&	-0.41	&	0.691   &	48	&	2.9	&	12	&   Polynomial	&	82 $\pm$ 16     \\
S0-17	&	15.9	&	Late	&	0.05	&	0.008	&	0.048   &	44	&	90	&	15	&	   Orbit	&	13.639 $\pm$ 0.090    \\
S0-18	&	14.9	&	Late	&	-0.12	&	-0.42	&	0.441   &	18	&	4.0	&	12	&   Polynomial	&	73 $\pm$ 15     \\
S0-27	&	15.5	&	Late	&	0.15	&	0.55	&	0.566   &	12	&	5.6	&	12	&   Polynomial	&	52 $\pm$ 10     \\
S1-5	&	12.4	&	Late	&	0.32	&	-0.89	&	0.943   &	27	&	2.6	&	12	&   Polynomial	&	150 $\pm$ 30     \\
S1-6	&	15.4	&	Late	&	-0.96	&	0.74	&	1.217   &	19	&	7.7	&	10	&   Polynomial	&	163 $\pm$ 33     \\
S1-10	&	14.7	&	Late	&	-1.10	&	-0.02	&	1.099   &	22	&	5.2	&	12	&   Polynomial	&	116 $\pm$ 23     \\
S1-13	&	14.0	&	Late	&	-1.14	&	-0.97	&	1.501   &	7	&	5.2	&	12	&   Polynomial	&	[158]     \\
S1-15	&	14.0	&	Late	&	-1.36	&	0.49	&	1.443   &	23	&	3.9	&	11	&   Polynomial	&	[152]     \\
S1-31	&	15.6	&	Late	&	-0.99	&	0.54	&	1.125   &	16	&	7.6	&	11	&   Polynomial	&   [118]	 \\
\enddata
\tablenotetext{a}{From Sgr A*.}
\tablecomments{Col 1: star name, Col 2: magnitude in $K'$, Col 3-5: projected distance from SgrA*, Col 6: total radial velocity points used in analysis, Col 7: median radial velocity uncertainty, Col 8: baseline of radial velocity measurements, Col 9: method for subtracting long-term RV trend, Col 10: semi-major axes estimates (values in brackets come from averaging comparably large separations).}
\end{deluxetable*}


%% file: polynomial_table_09-30-22.tex
\begin{deluxetable*}{lrrcc}
\tablecolumns{5}
\tablewidth{0pc} 
\tablecaption{Polynomial Fit Results \label{tab:polynomial_table_test}}
\tablehead{
\colhead{Star}	&
\colhead{$t_{0}$}   &
\colhead{$v_{z0}$}	&
\colhead{$a_{z}$\tna}	&
\colhead{$\chi^{2}_{red}$}  \\
\colhead{}	&
\colhead{(Epoch)}	&
\colhead{(\kms)}	&
\colhead{(\kms yr$^{-1}$)}	&
\colhead{}  
}
\startdata
S0-7	&   2014.4978   &	105.3	$\pm 5.9	$		&	[6.9]				&	3.3		\\
S0-9	&   2014.3799   &	114.5	$\pm 4.8	$		&	[4.1]				&	2.1		\\
S0-11	&   2014.2696   &	-22.2	$\pm 4.6	$		&	[4.7]				&	1.5		\\
S0-14	&   2013.8235   &	-31.3	$\pm 2.8	$		&	[2.2]				&	1.0		\\
S0-15	&   2013.6498   &	-552.5	$\pm 4.8	$		&	[4.1]				&	2.9		\\
S0-31	&   2012.5771   &	-119	$\pm 11	$		&	[9.3]				&	1.7		\\
S1-8	&   2012.2612   &	-112.0	$\pm 7.2	$		&	[5.9]				&	1.9		\\
S0-6	&   2013.6093   &	90.38	$\pm 0.42	$		&	$0.83 \pm 0.12$	&	1.7			\\
S0-12	&   2013.9746   &	-39.26	$\pm 0.48	$		&	[0.44]				&	2.0		\\
S0-13	&   2013.9545   &	-45.11	$\pm 0.41	$		&	[0.35]				&	2.2		\\
S0-18	&   2014.4520   &	-289.3	$\pm 1.0	$		&	[1.0]				&	3.5		\\
S0-27	&   2013.8851   &	-121.2	$\pm 1.5	$		&	[1.3]				&	4.3		\\
S1-5	&   2014.4119   &	11.20	$\pm 0.50	$		&	[0.38]				&	1.9		\\
S1-6	&   2013.8535   &	-42.0	$\pm 1.5	$		&	[1.7]				&	1.1		\\
S1-10	&   2013.7375   &	-33.6	$\pm 1.0	$		&	[0.77]				&	1.4		\\
S1-13	&   2012.4453   &	-749.1	$\pm 1.6	$		&	[1.2]				&	6.8		\\
S1-15	&   2011.8983   &	-120.40	$\pm 0.82	$		&	[0.65]				&	2.1		\\
S1-31	&   2014.0749  	&	182.9	$\pm 1.6	$		&	[1.7]				&	1.9		\\
\enddata
\tablenotetext{a}{3$\sigma$ limits are given in brackets. The value is reported with its 1$\sigma$}
\tablecomments{Col 1: star name, Col 2: $t_0$ epoch from polynomial radial velocity fit, Col 3: constant velocity offset, Col 4: acceleration parameter, Col 5: $\chi^{2}_{red}$ of the fit.}
\end{deluxetable*}

%% file: S0-14_rvtable_example.tex
\begin{deluxetable*}{ccrrrcc}
\tablecolumns{7}
\tablewidth{0pc}
\tablecaption{Radial Velocities and Residuals \label{tab:S0-14_example_rv}}
\tablehead{
\colhead{Epoch}	&
\colhead{MJD}   &
\colhead{RV}	&
\colhead{RV $\sigma$}	&
\colhead{Residual}  &
\colhead{Source}    &
\colhead{Reference}\\
\colhead{(Year)}	&
\colhead{}	&
\colhead{(\kms)}	&
\colhead{(\kms)}	&
\colhead{(\kms)}  &
\colhead{}  &
\colhead{}
}
\startdata
2006.461 & 53904.50 & -81 & 23 & -50 & S0-14 \\
2006.497 & 53917.50 & -46 & 16 & -15 & S0-14 \\
2007.384 & 54241.50 & -4 & 20 & 26 & S0-14 \\
... & ... & ... & ... & ... & ... \\
\enddata
\tablecomments{A full electronic version will be published in the journal. Col 1: Julian Year, Col 2: modified Julian date, Col 3: radial velocity corrected to local standard of rest, Col 4: radial velocity uncertainty, Col 5: residual, Col 6: star, Col 7: if noted, radial velocity reported in following reference.}
\end{deluxetable*}

%% file: periodicity_result_withk.tex
\begin{deluxetable*}{llccccc}
\tablecolumns{7}
\tablewidth{0pc} 
\tablecaption{Companion Star Search Results\label{tab:periodicity_results}}
\tablehead{
\colhead{Star}	&
\colhead{Spectral}	&
\colhead{LS Amp}	&
\colhead{Monte Carlo}	&
\colhead{Bootstrap False Alarm}	&
\colhead{Has Detected}	&
\colhead{$K$ Limit}	\\
\colhead{}	&
\colhead{Type}	&
\colhead{Significance\tna (\%)}	&
\colhead{Significance\tna (\%)}	&
\colhead{Significance\tna (\%)}	&
\colhead{Companion}	&
\colhead{(\kms)}
}
\startdata
S0-1 & Early & 89.64 & 92.34 & 99.94 & No & 44.2 \\
S0-2 & Early & 69.30 & 72.00 & 80.32 & No & 8.5 \\
S0-3 & Early & 75.80 & 94.70 & 90.03 & No & 23.9 \\
S0-4 & Early & 82.73 & 92.50 & 98.66 & No & 56.4 \\
S0-5 & Early & 65.42 & 94.88 & 98.52 & No & 38.4 \\
S0-7 & Early & 96.11 & 93.80 & 92.38 & No & 62.9 \\
S0-8 & Early & 75.54 & 96.40 & 99.79 & No & 50.7 \\
S0-9 & Early & 85.06 & 95.90 & 92.38 & No & 41.5 \\
S0-11 & Early & 79.26 & 90.20 & 86.80 & No & 38.7 \\
S0-14 & Early & 82.03 & 98.60 & 70.99 & No & 15.7 \\
S0-15 & Early & 85.95 & 93.50 & 33.37 & No & 56.5 \\
S0-16 & Early & 90.52 & 67.00 & 99.27 & No & 82.4 \\
S0-19 & Early & 56.10 & 64.00 & 99.73 & No & 50.1 \\
S0-20 & Early & 67.17 & 96.60 & 52.88 & No & 90.4 \\
S0-31 & Early & 79.38 & 100.00 & 42.34 & No & 75.7 \\
S0-6 & Late & 76.83 & 97.18 & 44.18 & No & 3.1 \\
S0-12 & Late & 85.69 & 99.00 & 94.57 & No & 3.9 \\
S0-13 & Late & 84.13 & 90.50 & 86.34 & No & 3.4 \\
S0-17 & Late & 60.64 & 75.00 & 99.90 & No & 21.1 \\
S0-18 & Late & 99.57 & 97.80 & 73.03 & No & 21.1 \\
S0-27 & Late & 97.65 & 90.30 & 52.30 & No & 20.5 \\
S1-5 & Late & 90.88 & 92.70 & 4.47 & No & 4.75 \\
S1-6 & Late & 95.54 & 96.30 & 63.66 & No & 11.7 \\
S1-8 & Early & 96.92 & 90.40 & 29.82 & No & 59.2 \\
S1-10 & Late & 85.08 & 99.00 & 95.40 & No & 6.9 \\
S1-13 & Late & 99.74 & 100.00 & 45.61 & No & 41.4 \\
S1-15 & Late & 92.29 & 93.20 & 46.69 & No & 7.9 \\
S1-31 & Late & 88.97 & 82.50 & 51.89 & No & 17.7 \\
\enddata
\tablenotetext{a}{For the period with maximum Lomb-Scargle power}
\tablecomments{Col 1: star name, Col 2: spectral type, Col 3: Lomb-Scargle model amplitude significance, Col 4: Monte Carlo simulation significance, Col 5: bootstrap false alarm significance, Col 6: average significance from columns 3-5, Col 7: $K$ amplitude limit.}
\end{deluxetable*}

%% file: excluded_sample.tex
\startlongtable
\begin{deluxetable*}{lrcrrrl}
\tablecolumns{7}
\tablewidth{0pc} 
\tablecaption{Excluded Stars\label{tab:excluded_stars}}
\tablehead{
\colhead{Name}	&
\colhead{$K'$}	&
\colhead{Spectral}	&
\colhead{RA$\Delta$\tna}	&
\colhead{Dec$\Delta$\tna}	&
\colhead{R2D\tna}	&
\colhead{Exclusion}	\\
\colhead{}	&
\colhead{(mag)}	&
\colhead{Type}	&
\colhead{(\arcsec)}	&
\colhead{(\arcsec)}	&
\colhead{(\arcsec)}	&
\colhead{Reason}	
}
\startdata
S0-24 & 15.58 & Late & 0.20 & 0.09 & 0.22 & Confused \\
S0-26 & 15.20 & Early & 0.33 & 0.21 & 0.40 & Confused \\
S0-53 & 15.50 & Unknown & 0.35 & 0.20 & 0.40 & Confused \\
S0-28 & 15.45 & Late & -0.14 & -0.49 & 0.51 & Too Few RVs \\
S0-62 & 15.37 & Late & 0.16 & -0.54 & 0.57 & Confused \\
S0-29 & 15.45 & Late & 0.37 & -0.44 & 0.58 & Confused \\
S0-67 & 15.49 & Late & 0.25 & -0.54 & 0.59 & Confused \\
S0-33 & 15.95 & Unknown & 0.65 & -0.53 & 0.83 & Confused \\
S0-32 & 14.08 & Unknown & 0.32 & 0.79 & 0.85 & Foreground Star \\
S0-35 & 15.20 & Unknown & 0.02 & 0.88 & 0.88 & Confused \\
S1-3 & 12.09 & Early & 0.32 & 0.88 & 0.94 & Featureless \\
S1-26 & 15.41 & Late & -0.88 & 0.39 & 0.96 & Confused \\
S0-108 & 15.67 & Unknown & 0.45 & -0.90 & 1.01 &  Confused \\
S1-2 & 14.64 & Early & 0.08 & -1.02 & 1.02 &  Background Gas \\
S1-1 & 13.02 & Early & 1.04 & 0.03 & 1.04 &  Featureless \\
S1-27 & 15.80 & Early & -1.03 & 0.19 & 1.05 &  Confused \\
S1-29 & 15.26 & Early & 1.07 & 0.16 & 1.08 &  Confused \\
S1-4 & 12.43 & Early & 0.88 & -0.66 & 1.10 &  Featureless \\
S1-28 & 15.92 & Late & -0.37 & -1.05 & 1.12 &  Confused \\
irs16C & 9.91 & Early & 1.05 & 0.55 & 1.18 &  Wolf-Rayet \\
S1-32 & 15.15 & Late & -0.99 & -0.66 & 1.19 &  Confused \\
S1-7 & 15.73 & Late & -1.05 & -0.58 & 1.20 &  Confused \\
S1-85 & 15.50 & Unknown & 0.92 & -0.83 & 1.24 &  Confused \\
S1-33 & 14.94 & Early & -1.25 & -0.00 & 1.25 &  Background Gas \\
S1-86 & 15.30 & Unknown & 1.02 & 0.74 & 1.26 &  Confused \\
S1-12 & 13.41 & Early & -0.75 & -1.03 & 1.27 &  Featureless \\
S1-34 & 12.91 & Late & 0.87 & -0.99 & 1.32 &  Confused \\
S1-14 & 12.90 & Early & -1.32 & -0.37 & 1.37 &  Featureless \\
irs16SW & 9.98 & Early & 1.11 & -0.95 & 1.46 &  Wolf-Rayet \\
S1-40 & 15.63 & Unknown & -1.41 & -0.61 & 1.54 &  Confused \\
S1-21 & 13.21 & Early & -1.64 & 0.09 & 1.64 &  Featureless \\
S1-22 & 12.52 & Early & -1.57 & -0.52 & 1.65 &  Featureless \\
S1-51 & 14.91 & Unknown & -1.66 & -0.17 & 1.67 &  Confused \\
S1-45 & 15.19 & Unknown & -1.28 & 1.10 & 1.69 &  Confused \\
\enddata
\tablenotetext{a}{From Sgr A*.}

\end{deluxetable*}

%% file: mass_results_appendix.tex
\begin{deluxetable*}{lrrrrrr}
\tablecolumns{7}
\tablewidth{0pc}
\tablecaption{Companion Mass Limits\label{tab:mass_limits}}
\tablehead{
\colhead{Star} &
\colhead{Mean Mag} &
\colhead{Spectral} &
\colhead{Isochrone} &
\colhead{Upper Limit $M_{comp}$} &
\colhead{Upper Limit Mass} &
\colhead{Equal Mass}    \\
\colhead{}  &
\colhead{($K'$)}  &
\colhead{Type}  &
\colhead{Mass (\msun)}  &
\colhead{Mass (\msun)}  &
\colhead{Ratio} &
\colhead{Binary (\msun)}
}
\startdata
S1-5	 & 12.48	 & Late		& 1.2 		 & 	0.10 	&	 0.083	 &	1.2 	\\
S0-13	 & 13.24	 & Late		& 1.2 		 & 	0.10 	&	 0.083	 &	1.2 	\\
S0-15	 & 13.55	 & Early	& 20.2 		 & 	5.4 	&	 0.27	 &	16.7 	\\
S0-14	 & 13.57	 & Early 	& 20.0 		 & 	1.7 	&	 0.085	 &	16.3 	\\
S0-6	 & 13.95	 & Late 	& 1.2 		 & 	0.10 	&	 0.083	 &	1.2 	\\
S1-13	 & 13.96	 & Late 	& 1.2 		 & 	0.70 	&	 0.58	 &	1.2 	\\
S0-2	 & 14.02	 & Early 	& 17.5 		 & 	1.1 	&	 0.063	 &	14.2 	\\
S1-15	 & 14.04	 & Late 	& 1.2 		 & 	0.10 	&	 0.083	 &	1.2 	\\
S1-8	 & 14.08	 & Early 	& 17.0 		 & 	5.6 	&	 0.33	 &	13.9 	\\
S0-4	 & 14.15	 & Early 	& 16.7 		 & 	5.7 	&	 0.34	 &	13.7 	\\
S0-9	 & 14.24	 & Early 	& 16.3 		 & 	3.8 	&	 0.23	 &	13.4 	\\
S0-12	 & 14.27	 & Late 	& 1.2 		 & 	0.10 	&	 0.083	 &	1.2 	\\
S0-3	 & 14.53	 & Early 	& 14.9 		 & 	2.2 	&	 0.15	 &	12.1 	\\
S1-10	 & 14.66	 & Late 	& 1.2 		 & 	0.10 	&	 0.083	 &	11.7 	\\
S0-1	 & 14.68	 & Early 	& 13.9 		 & 	3.1 	&	 0.22	 &	1.2 	\\
S0-18	 & 14.92	 & Late 	& 1.2 		 & 	0.40 	&	 0.33	 &	11.4 	\\
S0-5	 & 14.97	 & Early 	& 12.6 		 & 	2.2 	&	 0.17	 &	1.2 	\\
S0-31	 & 15.03	 & Early 	& 12.3 		 & 	4.9 	&	 0.48	 &	10.3 	\\
S0-7	 & 15.12	 & Early 	& 11.9 		 & 	4.6 	&	 0.38	 &	10.1 	\\
S0-11	 & 15.13	 & Early 	& 11.9 		 & 	3.0 	&	 0.25	 &	9.7 	\\
S0-16	 & 15.3		 & Early 	& 11.0 		 & 	5.4 	&	 0.49	 &	9.7 	\\
S0-19	 & 15.36	 & Early 	& 10.9 		 & 	3.6 	&	 0.33	 &	9.0 	\\
S1-6	 & 15.38	 & Late 	& 1.2 		 & 	0.20 	&	 0.16	 &	8.9 	\\
S0-27	 & 15.54	 & Late 	& 1.2 		 & 	0.30 	&	 0.25	 &	1.2 	\\
S1-31	 & 15.59	 & Late 	& 1.2 		 & 	0.30 	&	 0.25	 &	1.2 	\\
S0-8	 & 15.79	 & Early 	& 9.0 		 & 	3.1 	&	 0.34	 &	1.2 	\\
S0-20	 & 15.85	 & Early 	& 8.8 		 & 	5.8 	&	 0.64	 &	7.4 	\\
S0-17	 & 15.85	 & Late 	& 1.2 		 & 	0.30 	&	 0.25	 &	7.2 	\\
\enddata
\tablecomments{Col 1: star name, Col 2: mean magnitude in $K'$, Col 3: spectral type, Col 4: mass from the isochrone, Col 5: upper limit on companion mass, Col 6: upper limit on the mass ratio, Col 7: mass of each component of an equal mass binary system that would emit same flux as the star's photometry.}
\end{deluxetable*}